\documentclass[aip, jcp, citeautoscript, amsmath, amssymb, reprint,
]{revtex4-2}
\usepackage[title]{appendix}
\usepackage{graphicx, tikz, orcidlink}
\usepackage{epstopdf}
\usepackage{ascmac}
\usepackage{ulem}
\usepackage{cancel}
\usepackage{here}
\usepackage{physics}
\usepackage{bm}
\usepackage{comment}

\usepackage{amsthm,color}
\newtheoremstyle{query}%
{}{}
{\color{red}}
{}
{\sffamily\bfseries}{:}{12pt}
{}
\theoremstyle{query}
\newtheorem{aq}{Author Query/Comment}

\newcommand{\baq}{\begin{aq}}
\newcommand{\eaq}{\end{aq}}

\begin{document}
\title{Discretized hierarchical equations of motion in mixed Liouville--Wigner space for two-dimensional vibrational spectroscopies of liquid water}
\date{Last updated: \today}

\author{Hideaki Takahashi\orcidlink{0000-0001-6465-2049}}
\author{Yoshitaka Tanimura\orcidlink{0000-0002-7913-054X}}
\email[Author to whom correspondence should be addressed: ]{tanimura.yoshitaka.5w@kyoto-u.jp}
\affiliation{Department of Chemistry, Graduate School of Science,
Kyoto University, Kyoto 606-8502, Japan}

\begin{abstract}
A model of a bulk water system describing the vibrational motion of intramolecular and intermolecular modes is constructed, enabling  analysis of its linear and nonlinear vibrational spectra, as well as the energy transfer processes between the vibrational modes. The model is described as a system of four interacting anharmonic oscillators nonlinearly coupled to their respective heat baths. To perform a rigorous numerical investigation of the non-Markovian and nonperturbative quantum dissipative dynamics of the model, we derive discretized hierarchical equations of motion in mixed Liouville--Wigner space (DHEOM-MLWS), with  Lagrange--Hermite mesh discretization being employed in the Liouville space of the intramolecular modes and  Lagrange--Hermite mesh discretization and Hermite discretization in the Wigner space of the intermolecular modes. One-dimensional infrared and Raman spectra and two-dimensional terahertz--infrared--visible and infrared--infrared--Raman spectra are computed as demonstrations of the quantum dissipative description provided by our model. 
\end{abstract}
\maketitle

\section{Introduction}
\label{sec:intro}
Water in the condensed phase is the mother of chemistry, providing an environment that makes a variety of chemical and biological reaction processes possible.\cite{Ball1999,OCSACR1999,bagchi_2013} One of the key challenges in investigating the physical chemistry aspects of water is the study of irreversible energy transfer processes arising from the high-frequency intramolecular modes that promote bond formation and bond breaking through complex hydrogen bonding,\cite{Tokmakoff2003H2O,Hynes2004} and the low-frequency intermolecular modes that realize irreversible nuclear motion.\cite{Ohmine_ChemRev93,Yagasaki_ACR42,Yagasaki_ARPC64} The interplay between these modes plays key roles in chemical reaction processes.\cite{TokmakoffNat2013}  Experimentally, such phenomena have been investigated by infrared (IR)\cite{Bertie96IRexp,IRexp2011} and third-order off-resonant Raman spectroscopies.\cite{Brooker1989Ramanexp,Raman2018exp} Because water exhibits very great inhomogeneity, laser measurements based on linear responses, such as one-dimensional (1D) IR and Raman spectroscopies, yield broadened spectral peaks, as a consequence of which  it is difficult to use these measurements to investigate the underlying mechanisms of complex molecular dynamics. 

Thus, ultrafast two-dimensional (2D) spectroscopies have been proposed in which experiments are conducted by varying the time intervals in laser pulse trains applied to the molecules.\cite{TM93JCP,mukamel1999principles,Cho2009,Hamm2011ConceptsAM} By plotting the  nonlinear optical response as a function of the time intervals between the laser pulses, it is possible to obtain detailed information about the molecular dynamics in condensed phases. \cite{Tokmakoff2016H2O,Tokmakoff2022,ElsaesserCPL2005,ElsaesserH2O,ElsaesserJPCA2007,HammTHz2012,Hamm2013PNAS,HammPerspH2O2017,grechko2018,Bonn2DTZIFvis2021}
However,  analysis of these 2D signals is difficult, because they arise from complex motions described by nonlinear response functions exhibiting complicated spectral profiles that are very sensitive to the  physical conditions and the  setup of the experimental system. Therefore, to enable this type of experiment to be conducted successfully, support from theoretical analysis based on a predictive model is crucial.\cite{mukamel1999principles,JansenChoShinji2DVPerspe2019}
Molecular dynamics (MD) simulations have proved to be powerful means for such spectral analysis, but only a few approaches are able to cover the wide frequency range needed to analyze the role of energy relaxation between the intramolecular and intermolecular interactions of water molecules.\cite{HT11JPCB,ImotXanteasSaitoJCP2013H2O,ChoH2OMD2014,PaesaniJCTC2014H2O} In particular, simulations of the intramolecular modes are difficult, because their motions have to be treated quantum mechanically.\cite{IT06JCP,ST11JPCA,Imoto_JCP135,PaesaniJCP2014qMD}
Because the computational time required for  2D vibrational spectra is about 1000 times as long as that  for linear spectra, full MD simulations of 2D vibrational spectroscopies have been conducted mainly for the intermolecular modes, where  classical descriptions work reasonably well,\cite{Shinji2DRaman2006,HammTHz2012,Hamm2013PNAS,hamm2014,HammPerspH2O2017,IHT14JCP,IHT16JPCL,HT08JCP,YagasakiSaitoJCP20082DIR} although full quantum MD simulations for liquid water have recently become possible.\cite{JianLiu2018H2OMP, Paesan2018H2OCMD,Althorpe2019CMD}

As a practical approach, the Brownian oscillator (BO) model has been employed for the analysis of both linear\cite{Kuehn2015JPCL, Kuehn2015JCP,Kuehn2016JCP} and nonlinear spectra.\cite{mukamel1999principles,TI09ACR} In this approach, the vibrational modes representing the spectroscopic properties of interest are described as functions of molecular coordinates, while the environmental molecular motions are described  using heat baths that exert thermal fluctuations and dissipation on the vibrational modes. While analysis based on 2D spectroscopy has shown that vibrational relaxation and dephasing are important mechanisms for characterizing molecular motions,\cite{YagasakiSaitoJCP20082DIR,YagasakiSaitoJCP2011Relax} the inclusion of linear--linear (LL) and square--linear (SL) non-Markovian system--bath (SB) interactions is significant,\cite{OT97PRE} in addition to anharmonic mode--mode interactions (the LL+SL BO model).\cite{IT07JPCA,TI09ACR} Then, to obtain an accurate numerical solution of the model in a nonperturbative regime, various hierarchical equations of motion (HEOM) approaches have been employed.\cite{T06JPSJ,T20JCP}

The multimode LL+SL BO model was employed for the analysis of classical MD results of 2D THz-Raman\cite{IIT15JCP} and 2D IR-Raman spectra\cite{IT16JCP} 
with the use of the classical hierarchical Fokker--Planck equations (CHFPE), which are the classical limit of the quantum hierarchical Fokker--Planck equations (QHFPE) in the Wigner space representation.\cite{TW91PRA} The  2D spectra obtained from the CHFPE accurately reproduce the 2D profile of the classical MD results. It should be noted that because the 2D spectral profiles are extremely sensitive to the essential features of the intermolecular and intramolecular motion, it is not possible to reproduce them satisfactorily without capturing the key features of the molecular vibrational motion that are necessary to reproduce the complex 2D profile from a simple model. 
Such a BO model analysis of the 2D IR-Raman spectrum predicted the presence of a cross peak representing the mode--mode interactions between  intramolecular OH stretching motion and  hydrogen-bonded (HB)-intermolecular vibrational (translational) motion,\cite{IT16JCP} which was later observed  experimentally.\cite{grechko2018,Bonn2DTZIFvis2021}

Although the classical MD results for 2D water spectra explain the dynamical properties of water reasonably well, the intramolecular motion of water inherently requires a quantum treatment, especially to account for the peak splitting of stretching modes in the 2D IR spectrum caused by  transition between the 0-1-0 and 0-1-2 vibrational levels.\cite{ST11JPCA}

To aid in the construction of a reliable model, it would be helpful to have as a reference either experimental or theoretical results for  2D spectra  covering the entire vibrational spectral region. However, at present, no such results are available, and to overcome this difficulty, we attempt here to extend the classically constructed model\cite{IIT15JCP,IT16JCP} to the quantum case and solve it accurately in the framework of open quantum dynamics theory. It is not certain that our extended model in its present form will be able to  reproduce the experimental spectra if they are obtained, but modifications of the model parameters or model interactions to take account experimental and quantum simulation results of 2D spectrum should be straightforward. Once such a model has been successfully established, it could be used as a convenient tool for analyzing the energy relaxation process of water molecules, even in the quantum regime. Moreover, the model itself could be used as a quantum heat bath describing a complex water environment. Thus, the purpose of this study is to provide a basis for further analysis of complex water dynamics.

The remainder of this paper is organized as follows. In Sec.~\ref{sec:model}, we present the model Hamiltonian and HEOM. 
We then introduce the discretized HEOM in a mixed Liouville--Wigner space in Sec.~\ref{sec:DHEOM-MLW}. Although there are four major modes of water, two modes are sufficient for the calculation of a linear (1D) spectrum with a combination band peak and a 2D spectrum for detection of anharmonic mode--mode coupling. 
Thus, in Sec.~\ref{sec:result}, we present the calculated results for 1D IR and 1D Raman, 2D THz--IR--Visible and 2D IR--IR--Raman spectra based on  two-mode calculations. Section~\ref{sec:conclusion} is devoted to concluding remarks. The computer codes for the DHEOM-MLWS used in the present calculations are provided as supplementary material.

\section{Multimode LL+SL BO model and HEOM}
\label{sec:model}
We consider a model that consists of four primary oscillator modes of liquid water representing (1)  intramolecular OH stretching (``stretching''), (2) intramolecular HOH bending (``bending''), (3) hydrogen-bonded (HB)-intermolecular librational (``librational''), and (4) HB-intermolecular translational (``translational'') motions. They are described by dimensionless vibrational coordinates $\vec{q}=(\dots,q_{s},\dots)$ with $s=1,\dots,4$  indexing the four vibrational modes.  Although the present model is constructed to simulate these four modes simultaneously, the calculation of the 2D spectrum takes about 1000 times longer than that of the 1D spectrum, and, in this study, we limit our analysis to various combinations of two-mode cases. The Hamiltonian of the $s$th mode is expressed as\cite{IIT15JCP,IT16JCP}
\begin{align}
  \hat{H}_{A}^{(s)}= \frac{\hat{p}_s^{2}}{2m_s} + U_s(\hat{q}_s),
\end{align}
where $m_s$ and ${\hat p_s}$ are the mass and momentum for the $s$th modes with $s=1,\dots,4$. The potential of the $s$th mode and the interaction between the modes $s$ and $s'$ are denoted by $U_s(\hat{q}_s)$ and ${U}_{ss'}(\hat{q}_s, \hat{q}_{s'})$, respectively.
Each mode is independently coupled to the optically inactive vibrational modes, which are regarded as a bath system.  This bath system is represented by an ensemble of harmonic oscillators. 
The total Hamiltonian with  counterterms is then expressed as\cite{IT07JPCA,TI09ACR,T06JPSJ,T20JCP,IIT15JCP,IT16JCP}
\begin{align}
  \hat{H}_\mathrm{tot}&= \sum_{s=1}^4 \qty( \hat{H}_{A}^{(s)} + \sum_{s>s'} \hat{U}_{ss'}\qty(\hat{q}_s, \hat{q}_{s'})) \\ \nonumber 
  &+ \sum_{j_s}\qty[\frac{\hat{p}_{j_s}^{2}}{2m_{j_s}}+\frac{m_{j_s}\omega_{j_s}^{2}}{2}\qty(\hat{x}_{j_s}-\alpha_{j_s} \hat{V}_s(\hat{ q}_s) )^2],
  \label{eq:h_total}
\end{align}
where the momentum, coordinate, mass, frequency, and coupling strength of the $j_s$th oscillator for the $s$th bath are given by ${p}_{j_s}$, ${x}_{j_{s}}$, $m_{j_{s}}$, $\omega _{{j_s}}$, and $\alpha _{j_s}$, respectively. 

The bath dynamics can be characterized by the spectral distribution function (SDF), defined as $J_s(\omega) \equiv \sum_{j_s} {\alpha^2_{j_s}} \delta (\omega-\omega_{j_s})/{2 m_s \omega_{j_s}}$, and the inverse temperature $\beta=1/k_{\mathrm{B}}T$, where $k_{\mathrm{B}}$ is  Boltzmann's constant and $T$ is the thermodynamic temperature. The thermal properties of the bath are then characterized by the symmetrized correlation function (SCF) and the relaxation function (RF), expressed as $C_s(t) = \hbar \int_{0}^{\infty} \dd \omega\, J_{s}(\omega) \coth({\beta\hbar\omega}/{2} ) \cos(\omega t)$
and $\Psi(t) = 2 \int_{0}^{\infty} \dd \omega\, {J_{s}(\omega)} \cos(\omega t)/\omega$, which are related by the fluctuation--dissipation theorem.\cite{T06JPSJ,TK89JPSJ1}

For the vibrational modes of water, the anharmonicity of the potential and the mode--mode interactions are weak. Thus, we assume the potential of the $s$th mode and the interaction between the $s$th and $s'$th modes as 
\begin{align}
  \hat U_s(\hat{q}_s)= \frac{1}{2} m_s \omega_s^2 \hat{q}_s^2 +\frac{1}{3!}g_{s^3}q_{s}^3
\end{align}
and
\begin{align}
  \hat{U}_{ss'}(\hat{q}_s, \hat{q}_{s'}) = \frac{1}{2}  \qty(g_{s^2s'}\hat{q}_s^2 \hat{q}_{s'} + g_{s{s'}^2} \hat{q}_s \hat{q}_{s'}^2 ),
\end{align}
respectively, where $\omega_s$ is the frequency of the $s$ th mode, and $g_{s^3}$, $g_{s^2s'}$, and $g_{s{s'}^2}$ represent the third-order anharmonicity.  The dipole operator and polarizability are defined as
\begin{equation}
  \hat{\mu} = \sum_{s} \mu_s \hat{q}_s + \sum_{s,s'} \mu_{ss'} \hat{q}_s \hat{q}_{s'}
  \label{eq:mu}
\end{equation}
and 
\begin{equation}
  \hat{\Pi} = \sum_{s} \Pi_s \hat{q}_s + \sum_{s,s'} \Pi_{ss'} \hat{q}_s \hat{q}_{s'},
  \label{Pi}
\end{equation}
respectively, where $\mu_s$ and $\mu_{ss'}$ are the linear and nonlinear elements, respectively, of the dipole moment, and $\Pi_s$ and $\Pi_{ss'}$ are those of the polarizability. Then, the vibrational modes interact through the mechanical anharmonic coupling (MAHC) described by ${g}_{s^2 s'}$ and ${g}_{ss'^2}$ and the electric anharmonic coupling (EAHC) described by ${\mu}_{ss'}$ and  $\Pi_{ss'}$.
 
The system part of the SB interactions is denoted as $\hat{V}_{s}({\hat{q}_s})$, which consists of linear--linear (LL) and square--linear (SL) SB interactions as follows:
\begin{align}
  \hat{V}_{s}(\hat{q}_s)\equiv \hat{V}^{(s)}_{\mathrm{LL}}\hat{q}_s+ \frac{1}{2} \hat{V}^{(s)}_{\mathrm{SL}}\hat{q}_s^{2},
\end{align}
with  coupling strengths $V^{(s)}_{\mathrm{LL}}$ and $V^{(s)}_{\mathrm{SL}}$.\cite{TS20JPSJ,KT02JCP1,KT04JCP,IT06JCP,ST11JPCA}

In this study, we assume the SDF in the Drude form as
\begin{equation}
  J_s(\omega)=\frac{m_s \zeta_s}{2\pi}\frac{ \gamma_{s}^{2}\omega}{\omega^{2}+\gamma_s^{2}}, 
  \label{eq:drude}
\end{equation}
where $\zeta_s$ is the SB coupling strength and $\gamma_s$ represents the inverse correlation time of the $s$th bath noise.
We then have 
\begin{equation}
  C_s(t) = \sum_{k=0}^{K_s} c_{k}^{(s)} \mathrm{e}^{-\nu_{k}^{(s)} |t|} + c_{\delta}^{(s)} \cdot 2 \delta(t)
  \label{eq:scorrf2}
\end{equation}
and
\begin{equation}
  \Psi_s(t) = r_{0}^{(s)} \mathrm{e}^{-\gamma_s |t|},
  \label{eq:relaxf2}
\end{equation}
where $\nu_{0}^{s} \equiv \gamma_s$, and $ c_{\delta}^{(s)}$ is a renormalization factor.\cite{IT06JCP,T06JPSJ,T20JCP}
The vibrational modes of the water are then described by the reduced density matrix elements in Liouville space, $\rho(t)= \rho (\{q_s, q_s'\}; t)$, with $s=1,\dots,4$.  The HEOM in its Liouville-space expression for this system is then given by\cite{T06JPSJ,T14JCP}
\begin{align}
  \label{eq:HEOM}
 & \pdv{t} \rho_{\vec{n}}(t) = \nonumber  \\
  &
  - \sum_{s=1}^4 \qty(\mathcal{\hat{L}}_{A}^{(s)} + \sum_{s>s'} \mathcal{\hat{L}}_{I}^{(s,s')} 
  + \sum_{k=0}^{K_s} n_{k}^{(s)} \nu_{k}^{(s)} + \hat{\Xi}^{(s)} ) \rho_{\vec{n}}(t)\nonumber\\[3pt]
   &-\sum_{s=1}^{4} \qty( \sum_{k=0}^{K_s}   \hat{\Phi}^{(s)} \rho_{\vec{n}+\vec{e_k^{(s)}}}(t)   
   + \sum_{k=0}^{K_s} n_{k}^{(s)} \hat{\Theta}_k^{(s)} \rho_{\vec{n}-\vec{e_k^{(s)}}}(t)),
\end{align}
where $\mathcal{\hat{L}}_{A}^{(s)} \equiv i (\hat{H}_{A}^{(s)})^{\times} /\hbar$, $\mathcal{\hat{L}}_{I}^{(s,s')} \equiv i \hat{U}_{ss'}^{\times} /\hbar$, $\hat{\Phi}^{(s)} =i \hat{V}_s^{\times} /\hbar $, 
\begin{align}
  \hat{\Theta}_0^{(s)} \equiv \frac{r_{0}^{(s)}\gamma_{s}}{2\hbar} \hat{V}_{s}^{\circ} + \frac{i c_{0}^{(s)}}{\hbar} \hat{V}_{s}^{\times},
  \label{eq:Theta0}
\end{align}
 $\hat{\Theta}_k^{(s)} \equiv c_k^{(s)}\hat \Phi^{(s)}$, and
\begin{align}
  \hat{\Xi}^{(s)} \equiv - \frac{c_{\delta}^{(s)}}{\hbar^{2}} \hat{V}_{s}^{\times} \hat{V}_{s}^{\times} + \frac{r_{0}^{(s)}}{2\hbar^{2}} \hat{V}_{s}^{\circ}\hat{V}_{s}^{\times}.
  \label{eq:Xi}
\end{align}
Here, we have introduced the superoperator notation $\hat{\mathcal{O}}^{\times} \rho \equiv  [\hat{\mathcal{O}}, \rho]$ and $\hat{\mathcal{O}}^{\circ} \rho \equiv \{\hat{\mathcal{O}}, \rho \}$ for any operator $\hat{\mathcal{O}}$.
The vector $\vec{n}_{s} = (n_{0}^{(s)}, n_{1}^{(s)}, \ldots, n_{K_s}^{(s)})$ consists of nonnegative integers,  and  $\vec{e}_{k}$ is the unit vector of the $k$th element.
The zeroth element $\rho_{\vec{0}, \ldots, \vec{0}}(t)=\rho(\{q_s, q_s'\}; t) $ corresponds to the original density element.  

\section{Discretized HEOM in Mixed Liouville--Wigner Space}
\label{sec:DHEOM-MLW}
The density matrix in the present model is a function of the eight-dimensional elements in  coordinate space, and the above HEOM cannot be solved easily using currently available computer resources. To reduce  computational costs, we discretize the HEOM, taking into account the characteristics of the vibrational modes of water. 

Hereinafter, we distinguish the intramolecular modes and the intermolecular modes by $s=1$ and 2, and $\bar s=3$ and 4, respectively, and express the reduced density matrix as $\rho(t)= \rho (\{q_s, q_s'\}; \{q_{\bar s}, q_{\bar s}'\}; t)$.

\subsection{Lagrange--Hermite mesh discretization for intramolecular modes}

In the simulation of high-frequency intramolecular modes ($\hbar \omega_s \gg k_B T$ with $s=1$ and 2), the quantum nature of the system, which is commonly described in terms of energy eigenstates, plays an important role.\cite{SkinnerCPL2004,IT06JCP,ST11JPCA} However, to accurately treat the highly excited vibrational states, which are only slightly populated, we have to deal with many low-temperature correction terms (LTCTs) in the HEOM formalism.

We then find that by using the coordinate-space representation, we do not have to treat the excited states with low population explicitly, and we can reduce the LTCT elements dramatically, while maintaining numerical accuracy.
Hence, we describe the system using the Liouville operator in  coordinate space and then employ the basis functions to discretize.
All of the system operators are then expressed in matrix form as the elements of the basis set. Basis function approaches such as the discrete variable representation (DVR) method\cite{light1985} and the Lagrange-mesh method (LMM)\cite{baye1986,baye2015} have been used to discretize the wave function in  coordinate space. While the DVR method has been applied to the HEOM for the investigation of the electron transfer problem of metallic surfaces\cite{erpenbeck2019}, we found that the Lagrange--Hermite mesh method (LHMM) is more efficient for the description of high-frequency vibrational modes.

The LHMM is a variational method utilizing Gaussian basis functions.  The wave function is then described as $\psi(q_s ; t ) = \sum_{j_s}^{N_s} \alpha_{j_s}(t) F_{j_s}(q_s)$, where
\begin{align}
  F_{j_s}(q_s) = (-1)^{N_s-{j_s}}(2h_{N_s})^{-1/2} {b_{s}}^{\frac{1}{2}} \frac{H_{N_s} (\frac{q}{b_{s}})}{q_s-q_{j_s}} \exp\left(-\frac{q^2}{2b_{s}^2} \right) \nonumber \\
  &
  \label{eq:basis}
\end{align}
and $N_s$ is the total number of the basis set for the $s$th mode, $H_{N_s}(x)$ is the $N_s$th Hermite polynomial, and $h_{N_s}$ is the squared norm of $H_{N_s}(x)$, which is given by $h_{N_s} = \sqrt{\pi 2^{N_s} {N_s}!}$. We use the scaling factor defined by $b_{s} \equiv \sqrt{{\hbar}/{m_s \omega_s}}$, where $\omega_s$ is the characteristic frequency of the oscillator. The total number of mesh points in the LHMM is then $N_s \times N_s$.

The reduced density matrix elements for intramolecular modes are now expressed in terms of the discretized Liouville space elements as
\begin{align}
  \rho(\qty{q_{s}, q_{s}'}; \qty{q_{\bar s}, q_{\bar s}'}; t)& \equiv  \rho( \qty{q_{\bar s}, q_{\bar s}'} ; t) \nonumber \\ 
  & \times 
  \prod_s \rho_{\qty{ j_s, j_{s}'}}(t)
  F_{j_s}(q_s) F_{j_{s}'}(q_{s}'),
\end{align}
where $\rho_{\qty{j_s, j_{s}'}}(t) \equiv  \alpha_{j_s}(t) \alpha_{j_{s}'}(t)$. 
The superoperators of the system in Liouville space for the $s$th mode,  $\hat{X}(q_s)$,   as  functions of $q_s$, are expressed as
\begin{align}
  \hat{X}^{\times/\circ }(q_s) \hat{\rho}(t) \rightarrow \bm{X}^{\times /\circ } \bm{\rho}(t),
  \label{eq:X}
\end{align}
where
\begin{align}
  \bm{X}^{\times}_{ij} = \qty[X(q_{j_l}) - X(q_{j_r})]\delta_{i,j}
\end{align}
and
\begin{align}
  \bm{X}^{\circ}_{ij} = \qty[X(q_{j_l}) + X(q_{j_r})]\delta_{i,j}.
\end{align}
Here, we define $j_l \equiv \lceil {i}/{N_s} \rceil$ and $j_r  \equiv 1 + (N_s-1) \bmod i$.  The superoperator for the commutator of the squared momentum operator $\qty(\hat{p}_{s}^2)^{\times}$ can  also be expressed as 
\begin{align}
  \qty(\hat{p}_{s}^2)^{\times} \hat{\rho}(t) \rightarrow \bm{p}_s^{2\times} \bm{\rho}(t),
  \label{Liouv}
\end{align}
where
\begin{align}
  &\qty[\bm{p}_s^{2\times}]_{ij} = \nonumber \\
  \times &\qty[\frac{2(-1)^{k - k'}}{(q_{k}-q_{k'})^2} (1-\delta_{k,k'}) + \frac{1}{3}(2N_s + 1 -q_{k}^2)\delta_{k,k'}] \delta_{l,l'} \notag \\
  &- \delta_{k,k'}\qty[\frac{2(-1)^{l - l'}}{(q_{l}-q_{l'})^2} (1-\delta_{l,l'}) + \frac{1}{3}(2N_s + 1 -q_{l}^2)\delta_{l,l'}].
\end{align}
Here, we define $k \equiv \lceil {i}/{N_s} \rceil$, $l  \equiv 1 + (N_s-1) \bmod i$, $k' \equiv \lceil {j}/{N_s} \rceil$, and $l'  \equiv 1 + (N_s-1) \bmod j$.

\subsection{Lagrange--Hermite mesh discretization and Hermite discretization for intermolecular modes}

In general, the equations of motion for the density operator in  Liouville space expressed as $\rho(\qty{q_s, q_s'})$ are simple, since the Liouvillian is a local operator in  coordinate space.  Nevertheless, a phase-space ($p$--$q$ space)-like description such as the Wigner representation has advantages, in particular for the investigation of intermolecular modes ($\bar s=3$ and 4), because such a distribution is a real function and is localized more or less centered at $p_{\bar s}=0$ and $q_{\bar s}=0$ for a system with weak anharmonicity, and this  allows better usage of computer memory. Moreover, for low-frequency vibrational modes, the heat baths are considered to be at a high temperature, and a semiclassical treatment can be justified.
Conversely, the density matrix element of $\{q_s, q_s'\}$ is a complex and nonlocalized function. 

Thus, we introduce the Wigner distribution function (WDF), defined as 
\begin{align}
  \bar{\rho}(\qty{ p_{\bar s}, q_{\bar s}}) &\equiv \int \frac{\dd r_3 \dd r_4}{(2 \pi \hbar)^2} \mathrm{e}^{-i (p_3 r_3 +p_4 r_4) }  \nonumber \\
 &\times \rho \qty(\qty{ q_{\bar s} +\frac{r_{\bar s}}{2}, q_{\bar s} -\frac{r_{\bar s}}{2} })
  \label{eq:mwdf}
\end{align}
for $\bar s=3$ and 4, while we fix the elements of $s=1$ and 2. 

For the expression of system operators in  Wigner space, we employ the star operator $\star $, which represents the Moyal product defined as 
\begin{align}
  \star &\equiv \exp \qty[{\sum_{\bar s=3}^4 \frac{\hbar}{2i} \qty(\underleftarrow{\partial }_{q_{\bar s}}\underrightarrow{\partial}_{p_{\bar s}}-\underrightarrow{\partial }_{q_{\bar s}}\underleftarrow{\partial}_{p_{\bar s}})}],
  \label{eq:moyal-star}
\end{align}
where we have introduced the differentiation operations from the left and right, which are defined by
\begin{align}
  \underrightarrow{\partial}_{x}f(x)=f(x)\underleftarrow{\partial }_{x}&\equiv \frac{\partial f(x)}{\partial x}.
\end{align}
In this expression, the commutator and anticommutator are then replaced by $\hat{\mathcal{O}}^{\times} \bar{\rho} \rightarrow \mathcal{O}\star \bar \rho -\bar \rho \star \mathcal{O} $ for any operator $\mathcal{O}$ in the Wigner representation.

The quantum Liouvillian in the Wigner representation is, for example, expressed as
\begin{align}
  \sum_{\bar s=3,4} \hat{\mathcal{L}}_A^{(\bar s)} \bar{\rho} + \mathcal{\hat{L}}_{I}^{(3,4)} \bar{\rho} &= \sum_{\bar s=3,4} \frac{p_{\bar s}}{m_{\bar s}}  \pdv{q_{\bar s}} \bar{\rho} \nonumber \\
  &+ \frac{i}{\hbar} [U_{\bar s}(q_{\bar s}) \star {\bar \rho} - \bar{\rho} \star U_{\bar s}(q_{\bar s}) ] \notag \\
  &+ \frac{i}{\hbar} \qty[{U}_{34}(q_{3},q_{4}) \star \bar{\rho} - \bar{\rho} \star U_{34}(q_{3},q_{4}) ].
  \label{eq:Liouvillian}
\end{align}
For the LL+SL SB interaction, the relaxation operators in the Wigner representation for the HEOM have been evaluated and presented in Refs.~\onlinecite{TS20JPSJ,KT02JCP1,KT04JCP,T15JCP}. 

Note that, as can be seen from Eq.~\eqref{eq:moyal-star}, higher-order terms can be omitted when the wavepackets in  momentum space are nearly Gaussian or the anharmonicity of the potential is weak. Moreover, in the case of the low-frequency intramolecular modes ($\hbar \omega_{\bar s} \le k_B T$ with $\bar s=3$ and 4), the quantum effects of the system and the bath are minor, and a semiclassical or even classical description of the system is reasonably accurate. Thus, here we omit higher-order Kramers--Moyal expansion terms and employ the classical Liouvillian expressed as
\begin{align}
  \sum_{\bar s=3,4} \hat{\mathcal{L}}_A^{(\bar s)} \bar{\rho} + \mathcal{\hat{L}}_{I}^{(3,4)} \bar{\rho} &= \sum_{\bar s=3,4} \frac{p_{\bar s}}{m_{\bar s}}  \pdv{q_{\bar s}} \bar{\rho} \nonumber \\
  &+  \qty(\pdv{U_{\bar s}(q_{\bar s})}{q_{\bar s}} + \pdv{U_{34}(q_{3},q_{4})}{q_{\bar s}})\pdv{p_{\bar s}}\bar{\rho}.
  \label{eq:Liouvilliancl}
\end{align}
The intramolecular and intermolecular anharmonic couplings are expressed as
\begin{align}
  \mathcal{\hat{L}}_{I}^{(s,{\bar s})}\bar{\rho} = \frac{i}{\hbar} \hat{U}^{\times}_{s{\bar s}}(\hat{q}_{s},q_{{\bar s}}) \bar{\rho} + \frac{1}{2} \pdv{\hat{U}^{\circ}_{s{\bar s}}(\hat{q}_{s},q_{\bar s})}{q_{{\bar s}}} \pdv{p_{\bar s}}\bar{\rho}.
\end{align}

In terms of the Lagrange--Hermite functions and Hermite functions, the WDF can be discretized as\cite{blackmore1985a}
\begin{align}
  \rho(\qty{q_{\bar s}, q_{\bar s}'};t) &\equiv \prod_{\bar s} \bar \rho^{\qty{ k_{\bar s},{j_{\bar s}}}}(t) \nonumber \\
 & \times \qty[\psi_{0}(p_{\bar s}) \psi_{k_{\bar s}}(p_{\bar s}) f_{j_{\bar s}}(q_{\bar s}) \mathrm{e}^{-\beta U_{{\bar s}}(q_{\bar s})/2}],
\end{align}
where
\begin{equation}
  \psi_{k_{\bar s}}^{({\bar s})}(p_{\bar s}) = \frac{1}{\sqrt{2^{k_{\bar s}} {k_{\bar s}}! a_{\bar s} \sqrt{\pi}}} H_{k_{\bar s}} \qty(\frac{p_{\bar s}}{a_{\bar s}}) \exp\qty(-\frac{p_{\bar s}^2}{2a_{\bar s}^2})
  \label{eq:basis_w_p}
\end{equation}
and
\begin{equation}
  f_{j_{\bar s}}^{({\bar s})}(q_{\bar s}) = (-1)^{N_{\bar s}-{j_{\bar s}}}(2h_{N_{\bar s}})^{-1/2} b_{\bar s}^{\frac{1}{2}} \frac{H_{N_{\bar s}}\qty(\frac{q_{\bar s}}{b_{\bar s}})}{q_{\bar s}-q_{j_{\bar s}}} \exp\qty(-\frac{q_{\bar s}^2}{2b_{\bar s}^2}).
  \label{eq:basis_w_q}
\end{equation}
Here, $K_{\bar s}$ and $N_{\bar s}$ are the total numbers of basis functions for  momentum and coordinate spaces, respectively. 
The function $H_k(x)$ is the $k$th Hermite polynomial, and $h_N$ is the squared norm of $H_N(x)$, which is given by $h_N = \sqrt{\pi 2^N N!}$. 
We use  scaling factors defined as $a_{\bar s} \equiv \sqrt{{2m_{\bar s}}/{\beta}}$ and $b_{\bar s} \equiv {1}/{\sqrt{\beta m_{\bar s} \omega^2}}$  for  momentum and coordinate spaces, respectively.

Then, any function of $q_{\bar s}$ for the $\bar s$th mode $Z^{(\bar s)}(q_{\bar s})$ can be expressed as  
\begin{align}
  Z^{(\bar s)}(q_{\bar s}) \bar{\rho}(t) \rightarrow  \bm{Z}^{(\bar s)} \bar{\bm{\rho}} (t),
  \label{Z}
\end{align}
where
\begin{align}
  \bm{Z}^{(\bar s)}_{i_{\bar s},j_{\bar s}} = Z(q_{j_{\bar s}}) \delta_{i_{\bar s},j_{\bar s}},
\end{align}
Hereinafter, we denote the tensor expression for the first derivative of a function $Z(q_{\bar s})$ with respect to $q_{\bar s}$ as 
\begin{align}
  \pdv{Z(q_{\bar s})}{q_{\bar s}} \rightarrow \bm{{}_{\partial q_{\bar s}} Z}.
\end{align}
The first derivative $\partial/\partial {q_{\bar s}}$ can be expressed as
\begin{align}
  \pdv{q_{\bar s}} \bar{\rho}(t) \rightarrow \bm{D}^{(\bar s)} \bar{\bm{\rho}} (t),
  \label{Z}
\end{align}
where
\begin{align}
  \bm{D}^{(\bar s)}_{i_{\bar s},j_{\bar s}} &\equiv (-1)^{i_{\bar s} - j_{\bar s}}\frac{1}{(q_{i_{\bar s}}-q_{j_{\bar s}})} (1-\delta_{i_{\bar s},j_{\bar s}}) \nonumber \\
 &  - \delta_{i_{\bar s},j_{\bar s}}\frac{\beta}{2} \left.\pdv{U_{\bar s}(q_{\bar s})}{q_{\bar s}}\right|_{q_{\bar s}=q_{i_{\bar s}}}.
\end{align}

Here, we introduce the creation and annihilation operators $\bar{b}_{\bar s}$ and $\bar{b}^{\dagger}_{\bar s}$, which act on $\bar{\rho}^{\qty{ k_{\bar s},{j_{\bar s}}}}$ to decrease and increase the number of $k_{\bar s}$ as follows:
\begin{align}
  \left\{
  \begin{array}{ll}
  \bar{b}_{\bar s} \bar{\rho}^{\qty{k_{\bar s},{j_{\bar s}}}} = \sqrt{k_{\bar s}} \bar{\rho}^{\qty{k_{\bar s}-1,{j_{\bar s}}}}, \\
  \bar{b}^{\dagger}_{\bar s} \bar{\rho}^{\qty{k_{\bar s},{j_{\bar s}}}} = \sqrt{k_{\bar s}+1} \bar{\rho}^{\qty{k_{\bar s}+1,{j_{\bar s}}}}.
  \end{array}
  \right.
\end{align}
With these operators, we can express the momentum operator as
\begin{align}
  p_{\bar s} \rightarrow \sqrt{\frac{m_{\bar s}}{\beta}}\qty(\bar{b}_{\bar s}+\bar{b}^{\dagger}_{\bar s}),
\end{align}
and
\begin{align}
  \left\{
  \begin{array}{ll}
  \underrightarrow{\partial}_{p_{\bar s}} \rightarrow - \sqrt{\frac{\beta}{m_{\bar s}}} \bar{b}_{\bar s}, \\
  \underleftarrow{\partial}_{p_{\bar s}} \rightarrow - \sqrt{\frac{\beta}{m_{\bar s}}} \bar{b}^{\dagger}_{\bar s}.
  \end{array}
  \right.
\end{align}

The LHM discretization and Hermite discretization are efficient because the WDF is localized in  $q_s$ and $p_s$ space for a system at high temperature (semiclassical) or in an overdamped condition, while the SB interaction operator $\hat{V}^{(s)}(q_{s})$ is still in  diagonal form as 
$V^{(s)}_{j_{s},j_{s}'}  \approx V^{(s)}(q_{j_{s}})\delta_{j_s,j_{s}'}$. Moreover, in general, the LMM is numerically stable in comparison with the finite-difference method.\cite{baye1986} Such features allow us to dramatically reduce the computational time required to integrate the HEOM, in particular as the size of a system increases.
Note that for an unbounded system and a rotationally invariant system, a Lagrange--Fourier mesh method\cite{schwartz1985} and a discrete Wigner function method\cite{IT21JCEL} are respectively more efficient.

\subsection{DHEOM-MLWS}

The elements of the discretized reduced density matrix are now expressed as $\bar{\rho}_{\qty{ j_s, j_{s}'} }^{\qty{ k_{\bar s} j_{\bar s}} } (t) \equiv  \rho_{\qty{ j_s, j_{s}'} }(t) \bar{\rho}^{\qty{ k_{\bar s},{j_{\bar s}}}} (t) $.  The reduced density operator is then expressed in  tensor form as
\begin{equation}
 \bar{\boldsymbol{\rho}} (t) \equiv \qty{\bar{\rho}_{j_1, j_{1}'; j_2, j_{2}' }^{k_{3} j_{3}; k_{4} j_{4}} (t) }.
\end{equation}
The discretized HEOM are then expressed as
\begin{align}
  \label{eq:DHEOM}
  \pdv{t} \bar{\bm{\rho}}_{\vec{n}}(t) &={} \nonumber \\
&  - \left({\bm{L}}^{\mathrm{intra}} + {\bm{L}}^{\mathrm{inter}}  + \sum_{\mathrm{intra}-\mathrm{inter}} {\bm{L}}_{I}^{(s,\bar{s})}  \right. \nonumber \\
  &+ \left . \sum_{\bar s=1}^{4} \qty[ \sum_{k=0}^{K_s} n_{k}^{(s)} \nu_{k}^{(s)} +  \bm{\Xi}^{(s)}]\right) \bar{\bm{\rho}}_{\vec{n}}(t) \notag \\[3pt]
  &-\sum_{s=1}^{2}\left( \sum_{k=0}^{K_s} \bm{\Phi}^{(s)} \bar{\bm{\rho}}_{\vec{n}+\vec{e}}(t)
  + \sum_{k=0}^{K_s} n_{k}^{(s)} \bm{\Theta}_k^{(s)} \bar{\bm{\rho}}_{\vec{n}-\vec{e}}(t) \right) \notag \\
  &-\sum_{\bar s=3}^{4} \left( \sum_{k=0}^{K_{\bar s}} \bm{\Phi}^{(\bar s)} \bar{\bm{\rho}}_{\vec{n}+\vec{e}}(t)
  + \sum_{k=0}^{K_{\bar s}} n_{k}^{(\bar s)} c_{k}^{(\bar s)} \bm{\Phi}_k^{(\bar s)} \bar{\bm{\rho}}_{\vec{n}-\vec{e}}(t) \right. \notag \\
  & +  n_0^{(\bar{s})} r_0^{(\bar{s})} \bm{\Lambda}_1^{(\bar{s})}\bar{\bm{\rho}}_{\vec{n}-\vec{e}}(t)   \nonumber \\
& \left.  -  \frac{n_0^{(\bar s)}(n_0^{(\bar{s})}-1)}{2} r_0^{(\bar{s})2} \bm{\Lambda}_2^{(\bar{s})} \bar{\bm{\rho}}_{\vec{n}-2\vec{e}}(t)\right),
\end{align}
where
\begin{align}
  \bm{L}^{\mathrm{intra}} = \sum_{s=1,2} \frac{i}{\hbar}\qty[\qty(\frac{\bm{p}_s^{2\times}}{2m_{s}}+\bm{U}_{s}^{\times})] + \frac{i}{\hbar} \bm{U}_{12}^{\times},
\end{align}
\begin{align}
  \bm{L}^{\mathrm{inter}}  = \sum_{\bar s=3,4} \frac{1}{\sqrt{\beta m}} \bm{D}^{(\bar s)} \qty(\bar{b}_{\bar s}+\bar{b}^{\dagger}_{\bar s})  + \qty(\bm{{}_{\partial_{q_{\bar{s}}}}U_{\bar s}} + \bm{{}_{\partial_{q_{\bar{s}}}}U_{34}})\bar{b}_{\bar s},
\end{align}
and 
\begin{align}
  \bm{L}_{I}^{(s,\bar{s})} &= \frac{i}{\hbar} \qty(\frac{1}{2}g_{s^2 \bar{s}}\bm{q}^{2\times}_{s}\bm{q}_{{\bar s}} + \frac{1}{2}g_{s\bar{s}^2}\bm{q}^{\times}_{s}\bm{q}_{{\bar s}}^{2}) \nonumber \\
 & + \frac{1}{2}\sqrt{\frac{\beta}{m}} \qty(\frac{1}{2}g_{s\bar{s}}\bm{q}^{2\circ}_{s} - g_{\bar{s}s}\bm{q}^{\circ}_{s}\bm{q}_{{\bar s}})\bar{b}_{\bar{s}}.
  \label{eq:DHEOMLI}
\end{align}
Here, $\sum_{\mathrm{intra-inter}}$ denotes  summation with respect to coupling between intramolecular and intermolecular mode and, to keep the notation simple, the unit matrix is not denoted,   The expressions for other auxiliary operators are presented in Appendix~\ref{sec:AOP}

\section{Numerical demonstrations}
\label{sec:result}

\begin{table*}[!tb]
  \caption{\label{tab:FitAll1} Parameter values of  multimode LL+SL BO model for (1)  stretching, (2) bending, (3) librational, and (4) translational modes. Here, we set the fundamental frequency as $\omega_{0} = 4000$~cm$^{-1}$.  The normalized parameters are defined as $\tilde{\zeta}_s \equiv (\omega_0/\omega_s)^2\zeta_s$, $\tilde{V}_{LL}^{(s)} \equiv (\omega_s/\omega_0)V_{LL}^{(s)}$, $\tilde{V}_{SL}^{(s)} \equiv V_{SL}^{(s)}$, $\tilde{g}_{s^3} \equiv (\omega_s/\omega_0)^3 g_{s^3}$,$\tilde{\mu}_{s} \equiv (\omega_0/\omega_s)\mu_{s}$, $\tilde{\mu}_{ss} \equiv (\omega_0/\omega_s)^2 \mu_{ss}$, $\tilde{\Pi}_{s} \equiv (\omega_0/\omega_s)\Pi_{s}$, and $\tilde{\Pi}_{ss} \equiv (\omega_0/\omega_s)^2 \Pi_{ss}$.}
\scalebox{0.9}{
\begin{tabular}{ccccccccccc}
  \hline \hline
s&    $\omega_s$ (cm$^{-1}$) & $\gamma_s/\omega_0$ & $\tilde{\zeta}_s$ & $\tilde{V}_{LL}^{(s)}$ & $\tilde{V}_{SL}^{(s)}$ & $\tilde{g}_{s^3}$ & $\tilde{\mu}_{s}$  & $\tilde{\mu}_{ss}$ & $\tilde{\Pi}_{s}$   & $\tilde{\Pi}_{ss}$\\
  \hline
1&   $3520$ & $5.0{\times}10^{-3}$ & $9$ & $ 0 $                     & $1.0$ & $-5.0{\times}10^{-1}$ & $ 3.3 $  & $1.2{\times}10^{-2}$ & $ 3.3 $ & $2.5{\times}10^{-2}$\\
2&  $1710$ & $2{\times}10^{-2}$ & $0.8$ & $ 0 $                   & $1.0$ & $-7{\times}10^{-1}$ & $ 1.8 $ & $0$ & $0.47$  & $-3.9{\times}10^{-2}$\\
  \hline
3&  $390 $ & $8.5{\times}10^{-2}$ & $8.3$ & $3.4{\times}10^{-3}$ & $1.0$ & $ 7{\times}10^{-3}$ & $ 21 $ & $0$  & $2.1$  & $-0.83$\\
4& $125 $ & $0.5$ & $2.8$ & $2.8{\times}10^{-3}$ & $1.0$ & $ 9.7{\times}10^{-2}$ & $ 26 $ & $2.1$  & $ 9.0 $  & $2.3$ \\
  \hline \hline\\
\end{tabular}
}
\end{table*}

\begin{table}[!tb]
\caption{\label{tab:FitAll2}Parameter values of  multimode LL+SL BO model for anharmonic mode--mode coupling and optical properties among (1)  stretching, (2) bending, (3) librational, and (4) translational modes.  The normalized parameters are defined as $\tilde{g}_{s^2s'}\equiv (\omega_0^3/\omega_s^2 \omega_s') g_{s^2s'}$, $\tilde{g}_{s{s'}^2}\equiv (\omega_0^3/\omega_s \omega_{s'}^2) g_{s{s'}^2}$,
$\tilde{\mu}_{ss'} \equiv (\omega_0^2/\omega_s \omega_s') \mu_{ss'}$ and $\tilde{\Pi}_{ss} \equiv  (\omega_0^2/\omega_s \omega_s') \Pi_{ss'}$.
}
\begin{tabular}{cccccccccccc}
  \hline \hline
  $\mathrm{s-s'}$ & $\tilde{g}_{s^2s'}$ & $\tilde{g}_{s{s'}^2}$ & $\tilde{\mu}_{ss'}$ & $\tilde{\Pi}_{ss'}$ \\
  \hline
  $\mathrm{1-2}$  & $0$ & $0.2$ & $2.0 \times 10^{-3}$ & $2.6 \times 10^{-3}$ \\
  $\mathrm{1-3}$  & $-3.9 \times 10^{-2}$ & $-3.9 \times 10^{-2}$ & $0.13$ & $0.19$ \\
  $\mathrm{1-4}$  & $-7.5 \times 10^{-2}$ & $-7.5 \times 10^{-2}$ & $0.43$ & $0.46$ \\
  $\mathrm{2-3}$  & $-1.5 \times 10^{-2}$ & $-1.5 \times 10^{-2}$ & $7.0$ & $4.0$ \\
  $\mathrm{2-4}$  & $-2.0 \times 10^{-2}$ & $-2.0 \times 10^{-2}$ & $3.1\times 10^{-2}$ & $3.1 \times 10^{-2}$ \\
  $\mathrm{3-4}$  & $0.23$ & $0.23$ & $7.8 \times 10^{-2}$ & $0.16$ \\
  \hline \hline\\
\end{tabular}
\end{table}

We now report the results of our numerical computations of the DHEOM-MLWS. 
We employed the parameter values of the multimode LL+SL BO model chosen to reproduce 2D IR--Raman spectra obtained from classical MD simulations\cite{IT16JCP} with the use of the POLI2VS force fields\cite{HT11JPCB}, which possess the essential capability of simulating both IR and Raman spectra. We then  modified the anharmonicity and bath parameters of the intramolecular modes to fit an experimentally obtained IR spectrum\cite{IRexp2011} and Raman spectra\cite{Brooker1989Ramanexp,Raman2018exp} that are consistent with the 1D spectra obtained from  quantum MD simulations with the POLI2VS force fields. \cite{JianLiu2018H2OMP} That is, the anharmonicity of the intramolecular modes were modified by factors of 10 and 5 for the OH stretching mode ($s=1$) and HOH bending mode ($s=2$), respectively, while the bath coupling strength and inverse noise correlation time $(\zeta_s, \gamma_s)$ were modified by factors of (1.5, 2) for $s=1$ and  (0.7, 1.1) for $s=2$ from the classical values presented in Ref.~\onlinecite{IT16JCP}. Moreover, we enhanced the optical properties $\tilde{\mu}_{ss'}$ and $\tilde{\Pi}_{ ss'}$ for the 1-4 and 2-3 mode--mode couplings by  factors of 2 and about 100, respectively, to reproduce their overtone peaks. The anharmonicity of the potential and the mode--mode coupling strength are listed in Tables~\ref{tab:FitAll1} and~\ref{tab:FitAll2}.  Here, we employ the normalized parameters  to compare the effect of anharmonicity with respect to the potential for each mode and mode--mode coupling.
The bath temperature was set to $T = 300~\mathrm{K}$ ($\beta \hbar \omega _{0} \approx 19.2$), with fundamental frequency $\omega_{0} = 4000~\mathrm{cm}^{-1}$, which was chosen as a frequency close to the OH stretching mode.

The numerical calculations carried out to integrate Eqs.~\eqref{eq:DHEOM}--\eqref{eq:DHEOMLI} were performed using the fourth-order low-storage Runge--Kutta (LSRK4) method.\cite{yan2017cjcp,IT19JCTC}  A truncated Pad{\'e} spectral decomposition presented in Appendix~\ref{sec:TPSD} was employed to obtain the expansion coefficients of the noise correlation functions.  To conduct numerical integrations, the hierarchy was truncated to satisfy the condition $\Delta_{\vec{n}} / \gamma_{\vec{n}} < \delta_{\mathrm{tol}}$,\cite{IT19JCTC} where $\Delta_{\mathrm{tol}}$ is the tolerance of the truncation, with $\gamma_{\vec{n}} = \sum_{s}\sum_{k=0}^{K_s} n_{k}^{(s)}$ and
\begin{equation}
  \Delta_{\vec{n}} = \prod_{s=1}^{4}\prod_{k}^{K_s} \frac{1}{(n_k^{(s)}!)^{0.05}}\qty(\frac{|c_k^{(s)}|}{\nu_k^{(s)}}).
\end{equation}  

By adjusting the number of basis functions, we can calculate various physical quantities with any desired accuracy. The number of basis functions used in the calculation in the Liouville space [$F_j(Q)$ in Eq.~\eqref{eq:basis}] for both (1) stretching and (2) bending modes were $N_{s=2}=7$, and those in  Wigner space [$f_j(Q)$ and $\psi_k(p)$ in Eqs.~\eqref{eq:basis_w_p}, and~\eqref{eq:basis_w_q}] for (3)  stretching and (4) librational modes were (3) $N_{s'=3}=24$ and $K_{s'=3}=12$ and (4) $N_{s'=4}=24$ and $K_{s'=4}=8$. 

 We calculated 1D IR and 1D Raman spectra, and 2D THz--IR--Visible (2D TIV) and 2D IR--IR--Raman (2D IIR) spectra, defined by the first-order and second-order response functions expressed in terms of the two-body and three-body correlation functions of optical observables, respectively, according to the procedure explained in Appendix~\ref{sec:2dvs}. To efficiently obtain a 2D spectral profile utilizing  a small number of data points, we employed the estimation of signal parameters via rotational invariance techniques (ESPRIT) described in Appendix~\ref{sec:ESPRIT}.

The computation time with a multithreaded Fortran code of DHEOM-MLWS using a personal computer widely available today was approximately $2$~h for the 2D TIV spectrum.

\subsection{Linear response: 1D IR and 1D Raman spectra}
\begin{figure}[htbp]
  \centering
  \includegraphics[keepaspectratio, scale=0.35]{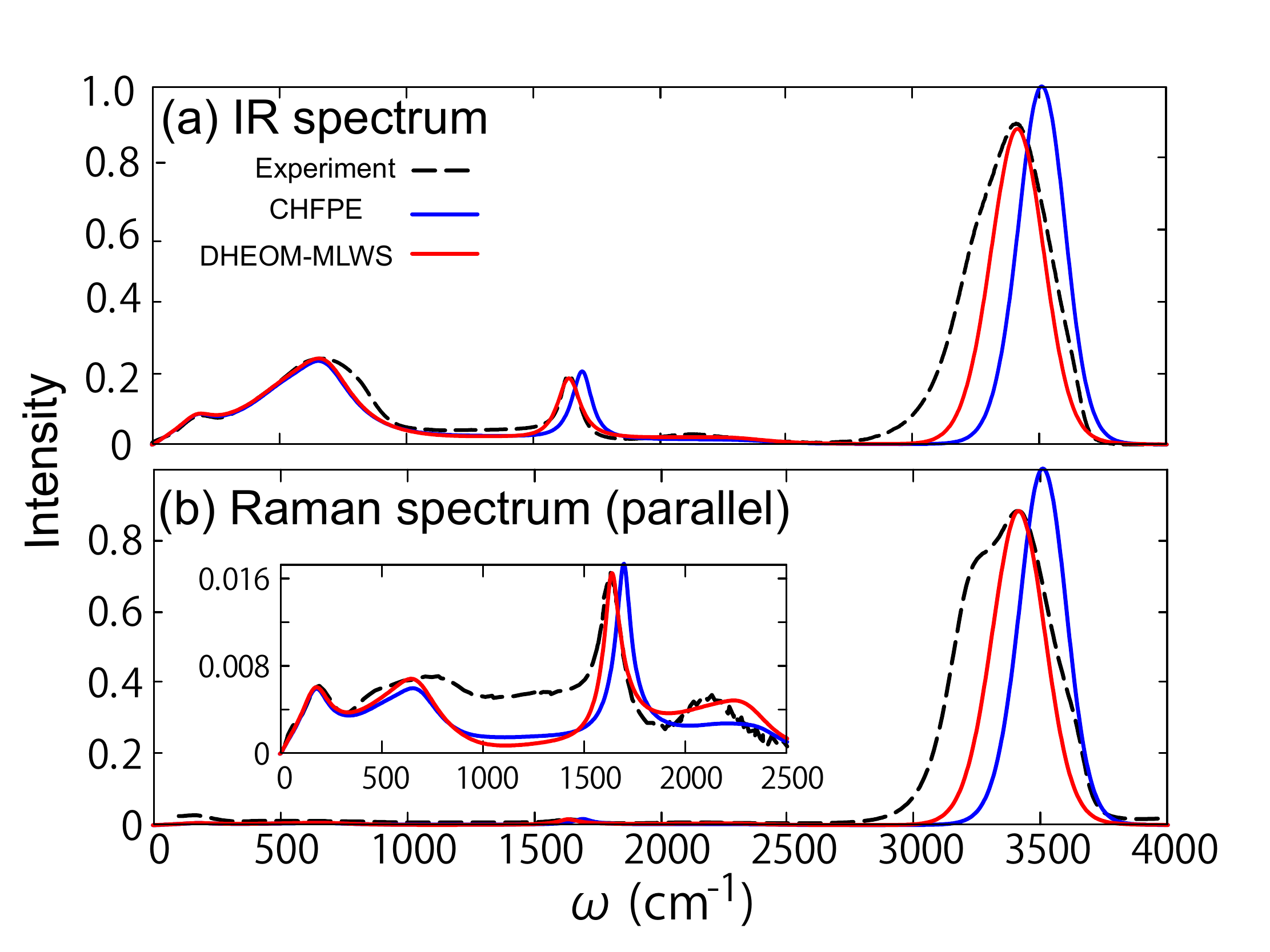}
  \caption{(a) 1D IR spectrum and (b) 1D parallel-polarized (VV) Raman spectrum of water calculated using classical and quantum HEOM approaches. The blue solid and red solid curves represent the classical CHFPE result and quantum DHEOM-MLWS result, respectively. The intensity of each spectrum is normalized with respect to the maximum peak intensity of the CHFPE results. The experimental IR\cite{IRexp2011} and Raman\cite{Brooker1989Ramanexp,Raman2018exp} data are also presented as  dashed curves for comparison.}
  \label{fgr:linear}
\end{figure}

In Figs.~\ref{fgr:linear}(a) and~\ref{fgr:linear}(b), we present the calculated 1D IR  and  1D Raman spectra, respectively, from  CHFPE and DHEOM-MLWS using the same parameter values of the present model. We obtained these spectra by combining the results from a single-mode model with $s=1$ and 2 and $\bar s=3$ and 4 and those from a two-mode model with the bending mode ($s=2$) and  stretching mode ($\bar s=4$), because the calculations for four modes are computationally expensive and because the effects of  mode--mode coupling are important only for the combination band among $s=2$ and $\bar s=4$.

In both the IR and Raman cases, the classical and quantum results agree for the low-frequency intermolecular modes. However, for the high-frequency intramolecular modes, the stretching and bending peaks in the classical case are blue-shifted owing to the quantum effects arise from the anharmonicity. Our DHEOM-MLWS results reproduce the weak bending--librational combination band at 2130~cm$^{-1}$ accurately, while the classical MD results\cite{HT11JPCB} and the classical BO results,\cite{IT16JCP} including the present CHFPE results, underestimate the peak position and peak intensity. 

Note that the present LL+SL BO model was constructed based on classical 2D IR-Raman simulations, but the force field used in the simulation (POLI2VS) was developed for quantum MD simulations. Because of this, although some modifications of the parameter values for the intramolecular modes are necessary, such an LL+SL BO model can predict a reasonably accurate vibrational spectrum when we conduct quantum HEOM calculations, as is also the case with quantum MD calculations using the POLI2VS force fields.\cite{JianLiu2018H2OMP} This indicates the possibility of constructing a quantum BO model from 2D spectra obtained from first-principles classical MD simulations, in which the nuclear motion of the molecules is classical.\cite{PaesaniJCTC2014H2O} 

\subsection{Nonlinear response: 2D THz--IR--visible and 2D IR--IR--Raman spectra}

\begin{figure}[htbp]
  \centering
  \includegraphics[keepaspectratio, scale=0.37]{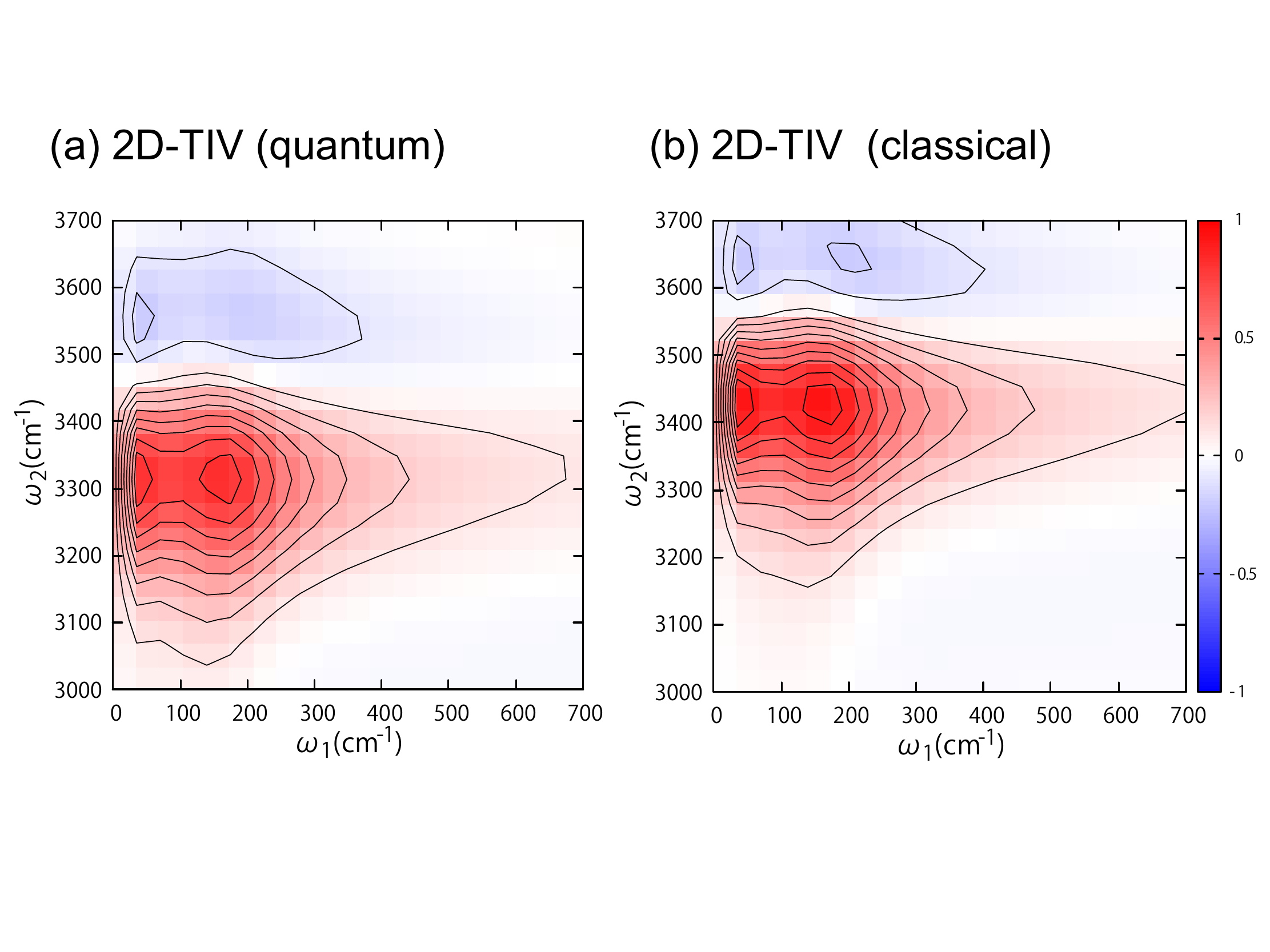}
  \caption{2D TIV spectra for the stretching--translational (1-4) modes calculated (a) with DHEOM-MLWS and (b) with CHFPE.  The spectral intensities are  normalized with respect to the absolute values of the spectral peak intensities of the classical result.}
  \label{fgr:stretch}
\end{figure}

\begin{figure}[htbp]
  \centering
  \includegraphics[keepaspectratio, scale=0.38]{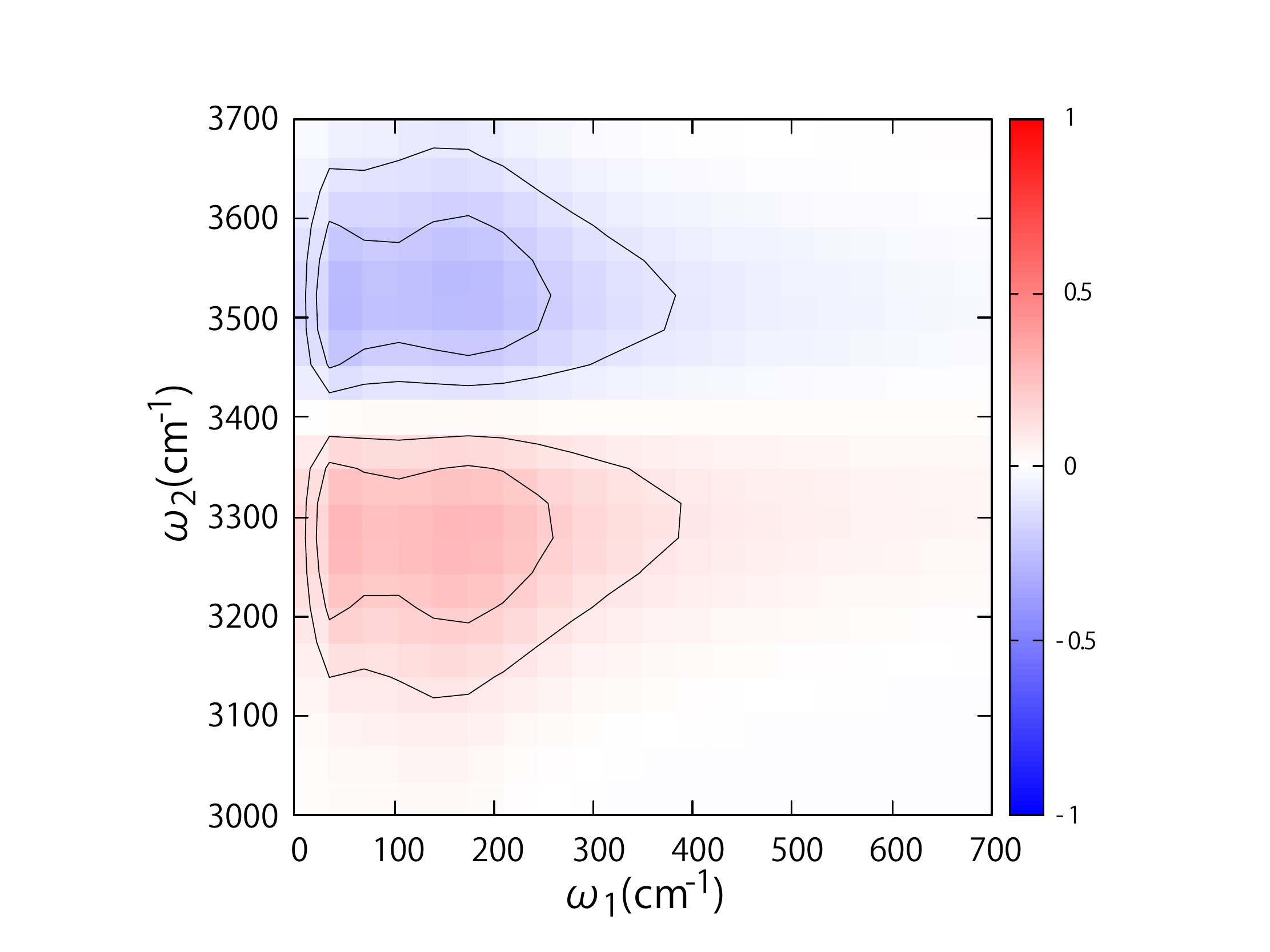}
  \caption{2D TIV spectrum for the stretching--translational (1-4) modes without the contribution from nonlinearity,  calculated with DHEOM-MLWS.  The spectral intensities are  normalized with respect to the absolute values of the peak intensities obtained from the classical simulation.}
  \label{fgr:stretch2}
\end{figure}

To elucidate how the effect of quantum dissipative dynamics is manifested in mode--mode coupling peaks, we next present numerical results for the 2D THz--IR--visible (2D TIV) spectrum\cite{grechko2018,Bonn2DTZIFvis2021}  and 2D IR--IR--Raman (2D IIR) spectrum (the observable part of the 2D TIV spectrum is equivalent to part of the 2D IIR spectrum presented in Ref.~\onlinecite{IT16JCP}). Note that in the 2D TIV and 2D IIR spectra expressed in terms of the three-body correlation functions of optical observables, the nondiagonal spectral peaks are not necessary to represent mode--mode coupling peaks as in the case of the third-order 2D IR spectrum expressed in terms of the four-body correlation functions of the dipole moment,\cite{mukamel1999principles,Cho2009,Hamm2011ConceptsAM} because the signal from the EAHC appears at a similar location to that from the MAHC.\cite{OT97CPL2,HammPerspH2O2017} 

Such 2D experiments have been conducted\cite{grechko2018,Bonn2DTZIFvis2021} on the basis of classical MD simulations,\cite{IT16JCP} but the results are not in good agreement with theoretical predictions, partly because of the classical description of the system. Hence, here we calculate and compare the 2D spectra for the classical and quantum cases using the same BO model, although the difference from the MD results may also be due to the limitation of the MD description. Note that although here we consider the 2D TIV and 2D IIR cases, 2D spectra computed from different pulse configurations such as 2D IR--Raman--IR and 2D Raman--IR--IR spectra exhibit similar profiles, because the difference between the IR and Raman spectra determined from the EAHC is minor in our calculations based on the BO model.

In Figs.~\ref{fgr:stretch}(a) and~\ref{fgr:stretch}(b), we compare 2D TIV results under the same conditions, calculated for the quantum (DHEOM-MLWS) and classical (CHFPE) cases, respectively. Characteristic features of anharmonicity and nonlinear polarizability on such 2D spectral profiles in a single-mode case and a two-mode case described by the BO model were elucidated in Ref.~\onlinecite{IT16JCP}. From that analysis, the negative peak [at $(\omega_1, \omega_2)=(150~\mathrm{cm}^{-1},3700~\mathrm{cm}^{-1})$ in the quantum case] arises only from the MAHC between the stretching--translational modes, whereas the positive peak [around $(\omega_1, \omega_2)=(150~\mathrm{cm}^{-1},3400~\mathrm{cm}^{-1})$ in the quantum case] arises from  contributions from MAHC and EAHC. This can be easily confirmed by comparing the same calculation without the nonlinear polarizability ($\Pi_{s,s'}=0$).  Figure~\ref{fgr:stretch2} reveals  a negative peak and a positive peak whose node lines are centered at the resonant frequency, whereas we observe only a positive peak at the resonant frequency in the pure EAHC case (see Ref.~\onlinecite{IT16JCP}). As this fictitious model analysis has demonstrated, we can easily identify the key dynamics of a liquid water system that determine the 2D spectral profiles obtained from experiments and complex MD simulations. This is because, to reproduce a complex 2D spectral profile accurately from a simple model, the model must capture the dynamical properties of the system correctly.\cite{T06JPSJ,TI09ACR,T20JCP} 

Compared with  previous classical MD and BO model calculations, in the present calculation we observe  peaks near $(\omega_1, \omega_2)=(0~\mathrm{cm}^{-1},3700~\mathrm{cm}^{-1})$ and $(\omega_1, \omega_2)=(0~\mathrm{cm}^{-1},3400~\mathrm{cm}^{-1})$  in both the classical and quantum cases.  Such peaks arise for a vibrational system strongly coupled to an Ohmic bath, as has been demonstrated from the analytical expressions for 1D and 2D spectra.\cite{OT97PRE,ST02CPL}  Although the BO model that we have  employed here is similar to that used to analyze classical MD results, such low-frequency peaks could not be observed in the previous studies, because their MD simulation period in the $t_1$ direction was too short.\cite{IT16JCP} 

\begin{figure}[htbp]
\centering\includegraphics[keepaspectratio, scale=0.38]{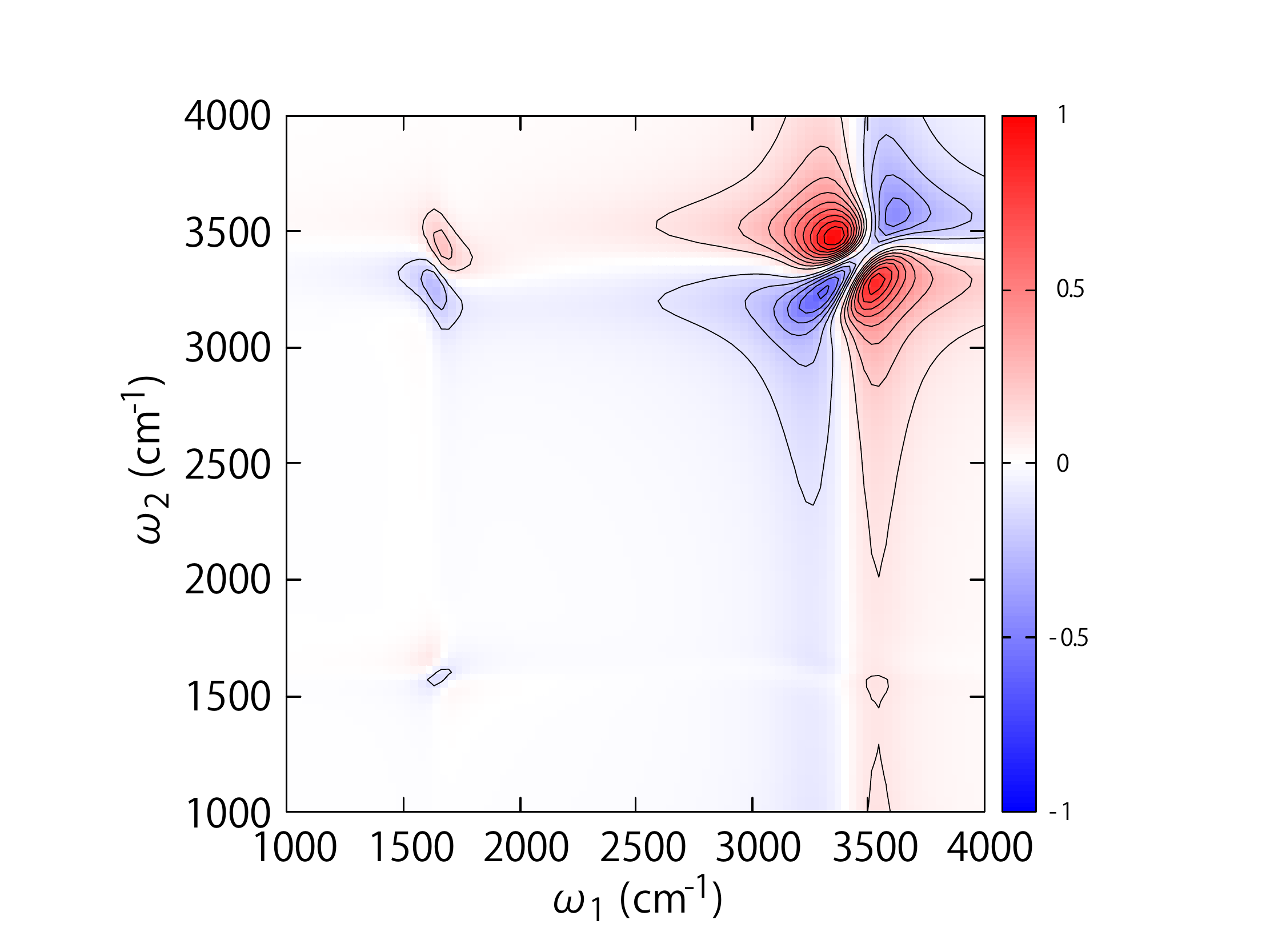}
 \caption{2D IIR spectrum for the stretching--bending (1-2) modes obtained from  DHEOM-MLWS. The spectral intensities are  normalized with respect to the absolute values of the spectral peak intensities.} 
\label{fgr:bending}
\end{figure}

We next demonstrate the description of the stretching--bending (1-2) modes. 
In Fig.~\ref{fgr:bending}, we depict the 2D IIR spectrum for the 1-2 modes. In this figure, there are both positive and negative peaks in the stretching, bending and their cross peak positions, indicating that these arise from the MAHC. This is because the EAHC contribution in the 1-2 modes is small, as indicated in Table~\ref{tab:FitAll2}. Here we observe the bending peak around $(\omega_1, \omega_2)=(1600~\mathrm{cm}^{-1},1600~\mathrm{cm}^{-1})$, but this peak overlaps with the bending-librational EHAC peak, as shown below, and cannot be identified. 

\begin{figure}[htbp]
  \centering
  \includegraphics[keepaspectratio, scale=0.4]{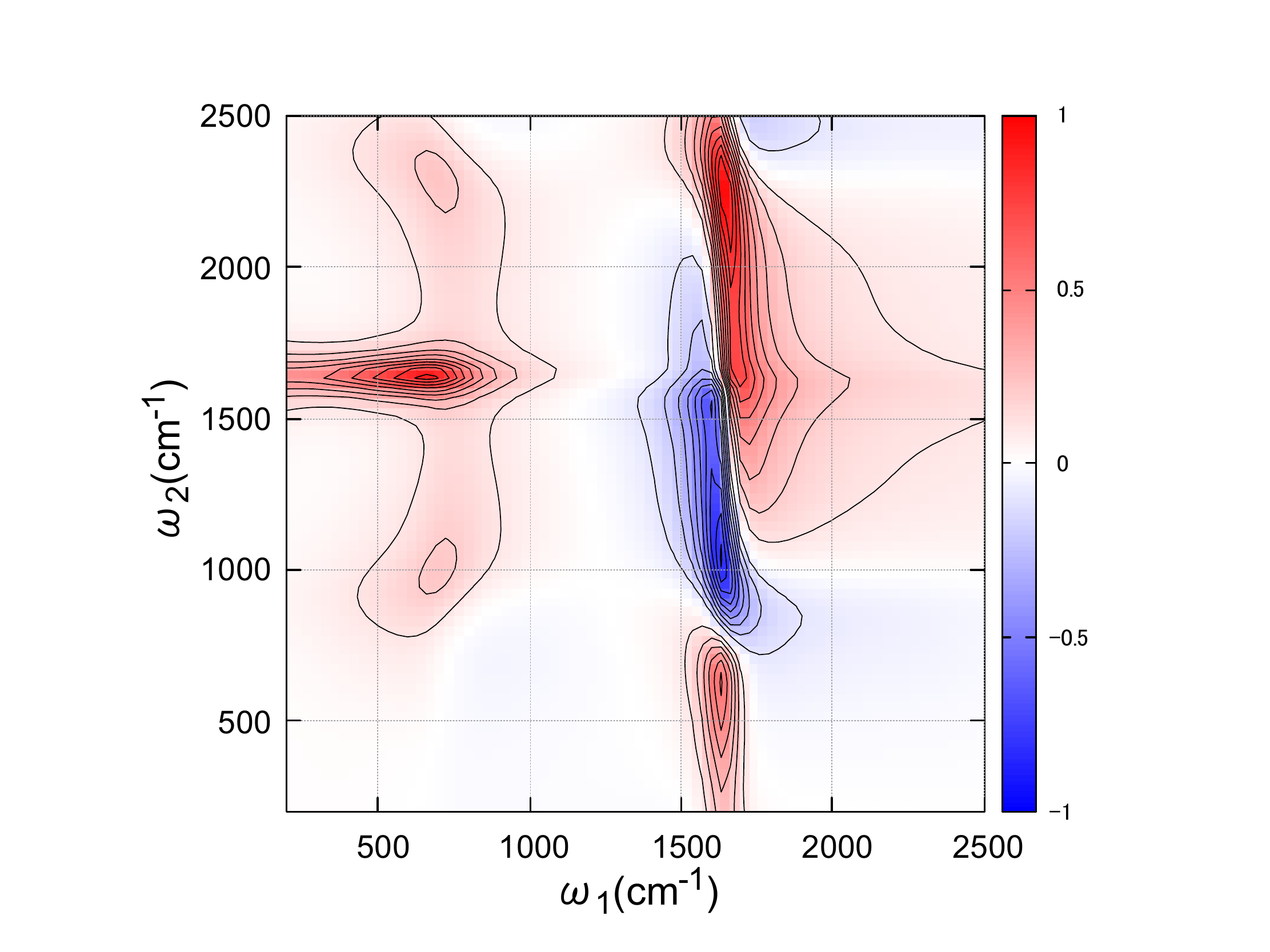}
  \caption{2D IIR spectrum for the bending--librational (2-3) modes obtained from  DHEOM-MLWS. The spectral intensities are  normalized with respect to the absolute values of the spectral peak intensities.} 
    \label{fgr:libration}
\end{figure}

In Fig.~\ref{fgr:libration}, we depict the 2D IIR spectrum calculated for the bending--librational (2-3) modes. As in the previous 1-4 case, because we perform an accurate quantum mechanical evaluation of the bending mode, the 2-3 coupling peaks around $(\omega_1, \omega_2)=(600~\mathrm{cm}^{-1},1600~\mathrm{cm}^{-1})$ and $(\omega_1, \omega_2)=(1600~\mathrm{cm}^{-1},600~\mathrm{cm}^{-1})$ are blue-shifted in comparison with the classical results.\cite{IT16JCP} The peak intensities of the 2-3 coupling peaks are much larger than the previous result\cite{IT16JCP} because we enhance the MAHC to reproduce the 2-3 combination band in the 1D IR spectra. The peaks that appear positively and negatively around $\omega_2 =1600 \mathrm{cm}^{-1}$ across $\omega_1 =1600 \mathrm{cm}^{-1}$ are caused by the EAHC of the 2-3 modes. Because we set $\Pi_{2,3}$ much larger value than classical case\cite{IT16JCP}, the bending peak displayed in Fig. ~\ref{fgr:bending} is completely covered by these EAHC peaks.

\section{Conclusion}
\label{sec:conclusion}
We have developed a model to analyze 1D and 2D vibrational spectra for both intramolecular and intermolecular vibrational modes involving all of their mode--mode interactions, taking into account the effects of energy relaxation and vibrational dephasing. To compute 2D signals from the model system, it is important to adopt a quantum-mechanically consistent treatment of the system and bath, in particular for the intramolecular modes, because the quantum entanglement between  system and bath plays an essential role. Thus, we adapted the HEOM formalism here,  enabling us to perform rigorous numerical calculations of linear and nonlinear spectra. Because integrating the HEOM for a multimode system is computationally expensive, we developed the DHEOM-MLWS approach to maintain the accuracy of the numerical calculation.

The description of the multimode LL+SL BO model with the use of the DHEOM-MLWS was investigated by calculating 1D and 2D spectra. From calculations of linear and nonlinear spectra, we obtained accurate predictions of the positions of the frequency stretching and bending peaks, for which the classical results are red-shifted.

The parameter values of our model were first chosen  by solving the classical HEOM to fit the classical MD results for  1D and 2D spectra,\cite{IT16JCP} with  the anharmonicity of the potentials being modified using the experimentally obtained 1D spectrum. We found that by using the POLI2VS force field in classical MD simulations, we could obtain a reasonable parameter value set for quantum HEOM simulation. This is because the POLI2VS force field was  developed for  quantum MD simulations,\cite{JianLiu2018H2OMP} and the complexity of molecular interactions, which is important in describing  water spectra, is not directly related to the issue of quantum effects.
This also indicates that, even using first-principles MD results,\cite{PaesaniJCTC2014H2O} in which the nuclear motion of the molecules is classical, we may construct a quantum SB model that includes complex anharmonic and bath interactions.

To reproduce a 2D spectral profile, the model must capture the dynamical properties of the vibrational motions correctly. Taking advantage of the low computational cost and simplicity of the model, we can easily examine, for example, the effects of higher-order anharmonicity on 2D spectra. The ability of the quantum mechanical model to calculate  2D spectra provides the possibility of directly analyzing experimentally obtained spectra. Once the model has been fully established, we can use it to  investigate energy and excitation transfer processes in  liquid water. Moreover, we can employ the LL+SL BO model as a heat bath to study the spectra and energy relaxation of liquids containing ions.

Extensions of the present model, for example, to describe symmetric and antisymmetric OH stretching modes separately and to employ the SDF to include the effects of optically inactive modes, are also possible.\cite{UT20JCTC} The present model with the HEOM approach provides a platform for analyzing novel experimental and simulation results.  We leave such extensions to future studies, depending on progress in experimental and simulation techniques. 

\section*{Supplementary material}
See the supplementary material for the computer codes for the DHEOM-MLWS used in the present calculations.

\begin{acknowledgments}
The authors are grateful to Shinji Saito and Keisuke Tominaga for helpful discussions. Y.T. was supported by JSPS KAKENHI (Grant No. B21H01884). H.T. is supported by JST SPRING (Grant No. JPMJSP2110).
\end{acknowledgments}

\section*{Author declarations}

\subsection*{Conflict of Interest}
The authors have no conflicts to disclose.

\section*{Data availability}
The data that support the findings of this study are available from the corresponding author upon reasonable request.

\appendix
\section{Auxiliary operators in DHEOM-MLWS}
\label{sec:AOP}
In this appendix, we present the explicit form of the auxiliary operators in the DHEOM-MLWS.
For intramolecular modes $s=1$ and 2, these are given by 
\begin{align}
  \bm{\Xi}^{(s)} \equiv - \frac{c_{\delta}^{(s)}}{\hbar^{2}} \bm{V}_{s}^{\times} \bm{V}_{s}^{\times} + \frac{r_{0}^{(s)}}{2\hbar^{2}} \bm{V}_{s}^{\circ}\bm{V}_{s}^{\times},
\end{align}
\begin{align}
  \bm{\Phi}^{(s)} = \frac{i}{\hbar} \bm{V}_{s}^{\times},
\end{align}
and
\begin{align}
  \label{eq:Theta0}
  \bm{\Theta}_k^{(s)} \equiv
  \left\{
  \begin{array}{ll}
  \frac{r_{0}^{(s)}\gamma_{s}}{2\hbar} \bm{V}_{s}^{\circ} + \frac{i c_{0}^{(s)}}{\hbar} \bm{V}_{s}^{\times} & (k=0),\\
  \frac{i c_{k}^{(s)}}{\hbar} \bm{V}_{s}^{\times} & (k>0).
  \end{array}
  \right.
\end{align}
For intermolecular modes $\bar s=3$ and 4, we have
\begin{align}
  \bm{\Xi}^{(\bar s)} \equiv -c_{\delta}^s\frac{\beta}{m_s} \bm{{}_{\partial q_{\bar s}}V}_{\bar s}^2 {\bar{b}_{\bar s}}^{2},
  \label{eq:XIint}
\end{align}
\begin{align}
  \bm{\Phi}^{(\bar s)} = \sqrt{\frac{\beta}{m_s}}\bm{{}_{\partial_{q_{\bar s}}}V}_{\bar s} \bar{b}_{\bar s},
\end{align}
\begin{align}
  \bm{\Lambda}_1^{(\bar s)} = -\sqrt{\frac{1}{\beta m_{\bar{s}}}}\bm{{}_{\partial_{q_{\bar s}}}V}_{\bar s} \qty(\bar{b}_{\bar s}+\bar{b}^{\dagger}_{\bar s}) + \frac{\hbar^2 V_{\mathrm{SL}}^{\bar s}}{4}\sqrt{\frac{\beta}{m_{\bar s}^3}}\bm{D}^{(\bar s)}\bar{b}_{\bar s},
\end{align}
and
\begin{align}
  \bm{\Lambda}_{2}^{\bar s} = \frac{\hbar^2 V_{\mathrm{SL}}^{\bar s}}{2}  \sqrt{\frac{\beta}{m_{\bar s}^3}} \bm{{}_{\partial_{q_{\bar s}}}V}_{\bar s}\bar{b}_{\bar s}.
  \label{eq:LambdaIint}
\end{align}
In the classical limit $\hbar \rightarrow 0$, Eqs. \eqref{eq:XIint}-\eqref{eq:LambdaIint} reduce to $\bm{\Xi}^{(\bar s)} \equiv 0$, 
\begin{align}
  \bm{\Phi}^{(\bar s)} = \sqrt{\frac{\beta}{m_s}}\bm{{}_{\partial_{q_{\bar s}}}V}_{\bar s} \bar{b}_{\bar s},
\end{align}
\begin{align}
  \bm{\Lambda}_1^{(\bar s)} = -\sqrt{\frac{1}{\beta m_{\bar{s}}}}\bm{{}_{\partial_{q_{\bar s}}}V}_{\bar s} \qty(\bar{b}_{\bar s}+\bar{b}^{\dagger}_{\bar s}) ,
\end{align}
and $\bm{\Lambda}_{2}^{(\bar s)} = 0$.

\section{Truncated Pad{\'e} spectral decomposition}
\label{sec:TPSD}
When quantum effects described by an SB model becomes important, we have to take into account many LTCTs involved in the HEOM formalism, which makes the integration of the HEOM computationally very expensive.  Thus, to reduce the number of  LTCTs, a Pad{\'e} spectral decomposition (PSD) scheme has been developed.\cite{YanPade2010} We can further reduce the number of hierarchical elements by incorporating  into the PSD the balanced truncation method (BTM), which was originally developed as a model order reduction (MOR).\cite{xu2013}  Here, we adapt the algorithm developed in Ref.~\onlinecite{ikeno2018} for the LTCTs and demonstrate the efficiency of the truncated PSD (TPSD) method. 

For a desired accuracy $\epsilon > 0$, we consider the condition
\begin{equation}
  ||C_{\mathrm{PSD}}(t)-C_{\mathrm{TPSD}}(t)|| < \epsilon,
  \label{TPSD}
\end{equation}
where $C_{\mathrm{PSD}}(t)$ is the SCF described by PSD, and $C_{\mathrm{TPSD}}(t)$ is the SCF described by PSD with BTM. 

\begin{figure}
  \centering
  \includegraphics[scale=0.35]{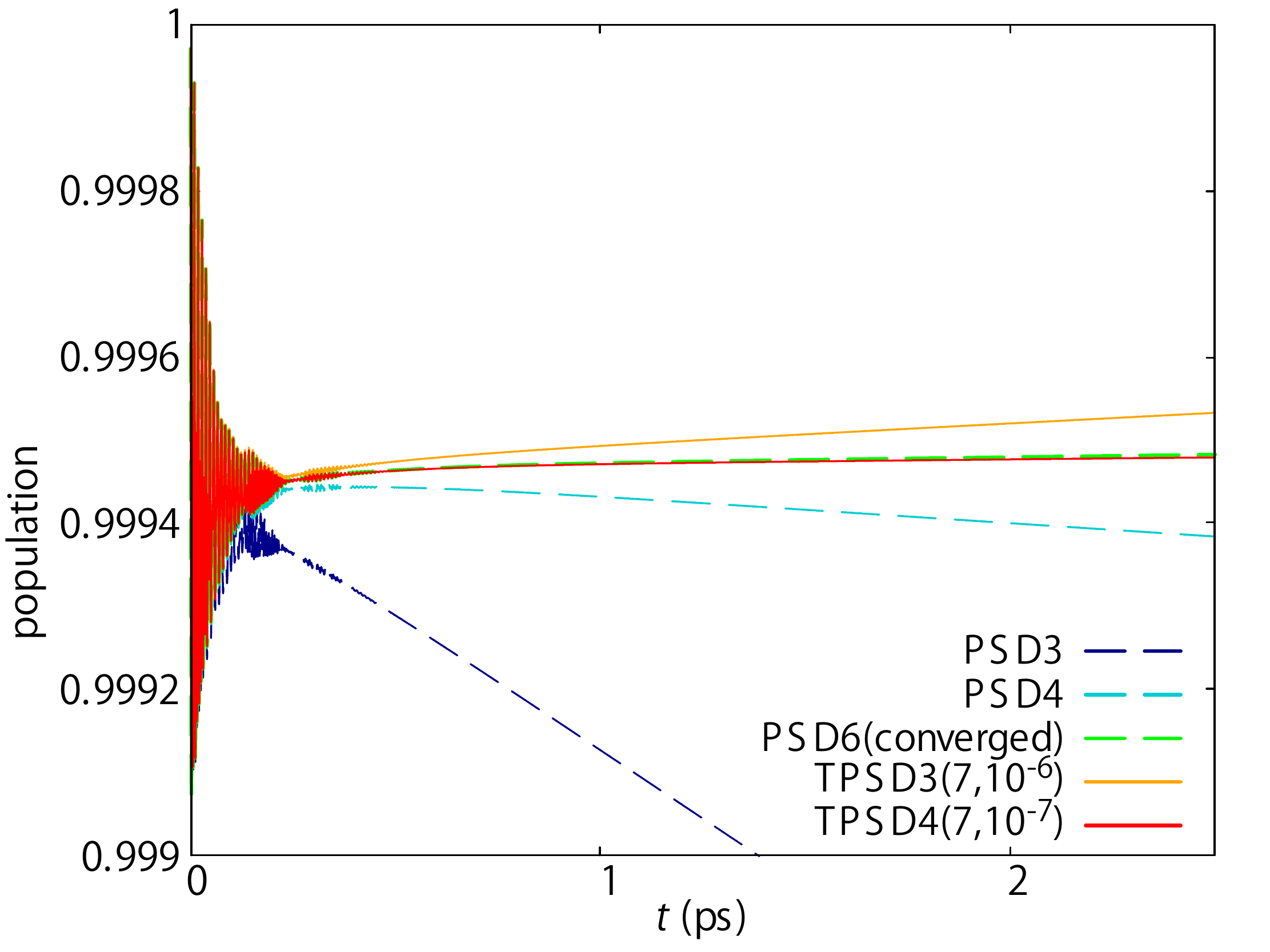}
  \caption{Time evolution of the ground-state population of the OH stretching mode calculated using HEOM with PSD and TPSD in the energy-eigenstate representation.  Here, PSD$n$ (dashed curves) represents the PSD with $n$ poles, and TPSD$n(M,\epsilon)$ (solid curves) represents the TPSD with $n$ poles utilizing the balanced truncation for PSD$M$ with tolerance $\epsilon$.}
  \label{fgr:tpsd}
\end{figure}

In Fig.~\ref{fgr:tpsd} we depict the time evolution of the ground-state population of the OH stretching mode ($s=1$) described using the LL+SL BO model with parameters values as listed in Table~\ref{tab:FitAll1}. We then employ the energy-eigenstate representation and  integrate the HEOM with PSD and with TPSD. As depicted in Fig.~\ref{fgr:tpsd}, the  results calculated with TPSD converge faster than those without TPSD, while the number of hierarchical terms is fewer in the case without PSD. The improvement becomes significant for a system strongly coupled to a heat bath at low temperatures. This approach is particularly beneficial when we deal with multiple heat baths. Note that this method can also  be combined with the NZ2 truncation method.\cite{fay2022b}

\section{Linear and nonlinear spectra}
\label{sec:2dvs}

Because the HEOM formalism is able to take accurate account of the quantum entanglement between  system and bath, it is possible to calculate linear nonlinear response functions. To compute a nonlinear spectrum in the HEOM approach, we express the response functions in terms of the time-propagation operator. For example, a 1D spectrum defined by first-order response functions is expressed in terms of the two-body correlation function as
\begin{align}R^{(1)}(t)=\frac{i}{\hbar }\mathrm{tr}\qty{\hat{A}\mathcal{G}(t)\hat{A}^{\times }\hat{\rho }^{\mathrm{eq}}},
  \label{eq:1DIR}
\end{align}
where we have employed the hyperoperator $^{\times }$ defined as $\hat{{A}}^{\times }\hat{{\hat \rho}}\equiv [\hat{{A}},{\hat \rho}]$ for the Liouville space representation and $\hat{A}^{\times} \hat{\rho} \equiv \hat{A} \star \hat{W} - \hat{W} \star \hat{A}$ for the Wigner space representation, $\mathcal{G}(t)$ is the Green's function of the system Hamiltonian without a laser interaction, and $\hat{\rho }_{\mathrm{eq}}$ is the equilibrium state. For 1D IR or 1D Raman calculations, we chose $\hat{A}=\hat \mu$ or $\hat{A}=\hat \Pi$ given in Eq.~\eqref{eq:mu} or~\eqref{Pi}.

Accordingly, the 2D spectrum defined by the second- and third-order response functions is expressed in terms of the three- and four-body correlation functions of optical observables as follows:\cite{T98CP,T06JPSJ}
\begin{align}  R^{(2)}(t_{2},t_{1})=\qty(\frac{i}{\hbar })^{2}\mathrm{tr}\qty{\hat{A}\mathcal{G}(t_{2})\hat{B}^{\times }\mathcal{G}(t_{1})\hat{C}^{\times }\hat{\rho }^{\mathrm{eq}}}
  \label{eq:ttr2}
\end{align}
and
\begin{equation}
  R^{(3)}(t_3,t_2,t_1) = \qty(\frac{i}{\hbar })^{3}\mathrm{tr}\qty{\hat{A}\mathcal{G}(t_{3})\hat{B}^{\times }\mathcal{G}(t_{2})\hat{C}^{\times }\mathcal{G}(t_{1})\hat{D}^{\times }\hat{\rho }^{\mathrm{eq}}},
  \label{eq:RFTIRI}
\end{equation}
where the operators $\hat{A}$, $\hat{B}$, $\hat{C}$,and  $\hat{D}$ are either the dipole moment $\hat \mu$ or the polarizability $\hat \Pi$.

The above equations represent the time evolution of the system under laser excitation. For example, Eq.~\eqref{eq:ttr2} can be interpreted as follows. The system is initially in the equilibrium state $\hat{\rho }^{\mathrm{eq}}$ and is then modified as a result of the first laser pulse via the dipole interaction by $\hat{C}$. It then propagates for time $t_{1}$ under $\mathcal{G}(t_{1})$. The system is next excited through the second laser pulse by $\hat{B}$ and propagates for time $t_{2}$ under $\mathcal{G}(t_{2})$. Finally, the expectation value of the polarizability at $t_{1}+t_{2}$ is generated through the laser pulses by $\hat{A}$.\cite{T06JPSJ,T20JCP}
The 2D THz--IR--visible signal can be computed from $R^{(2)}_\mathrm{TIV}(t_2, t_1)$ in Eq.~\eqref{eq:ttr2} with $\hat A=\hat \Pi $ and $\hat B=\hat C=\hat \mu$, and the 2D IR response $R^{(3)}_\mathrm{IR}(t_3, t_2,t_1)$  can be evaluated from Eq.~\eqref{eq:RFTIRI} with $\hat A= \hat B = \hat C= \hat D=\hat \mu$, respectively.  

From the second- and third-order response functions, the 2D TIV and 2D IR spectra, for example, are evaluated as
\begin{align} 
 & I_{\mathrm{TIV}}(\omega_1,\omega_2)  \nonumber \\  
  & \propto \int_{0}^{\infty} \mathrm{d} t_1 \int_{0}^{\infty} \mathrm{d} t_2 R_{\mathrm{TIV}}^{(2)}(t_2,t_1) \sin (\omega_1 t_1) \sin (\omega_2 t_2)
  \label{eq:2d}
\end{align}
and
\begin{align} 
  &I_{\mathrm{IR}}^{(3)}(\omega_1,t_2,\omega_3)  \nonumber \\  
  &   \propto \int_{0}^{\infty} \mathrm{d} t_1 \int_{0}^{\infty} \mathrm{d} t_3 R^{(3)}_\mathrm{IR}(t_3,t_2,t_1) \sin (\omega_1 t_1) \sin (\omega_3 t_3).
  \label{eq:3d}
\end{align}
This 2D sine-Fourier representation is more intuitive than the real part of the 2D-Fourier representation, since it can extract only absorptive components.

\section{Estimation of signal parameters via rotational invariance techniques}
\label{sec:ESPRIT}
Calculating 2D spectra is computationally expensive, because we have to repeat the dynamics calculations for different $t_1$ and $t_2$, or $t_1$, $t_2$, and $t_3$. Estimation of signal parameters via rotational invariance techniques (ESPRIT) leads to a dramatic reduction in the computational cost, because it allows us to find an optimal explored solution of a targeting signal as a linear combination of complex exponentials.
 
Thorough the use of  ESPRIT, the second-order response function as a function of $t_1$ is, for example, expressed as
\begin{equation}
  R(t_1,t_2) = \sum_{i=1}^{N} a_i(t_2) \mathrm{e}^{-b_i(t_2)t_1},
  \label{eq:ESPRIT_t}
\end{equation}
where $a_i(t_2)$ and $b_i(t_2)$ are  complex functions of $t_2$ that are chosen to optimize $R(t_1,t_2)$. Although the real part of $b_i(t_2)$ must be positive to avoid  divergence in the $t_1$ direction, we can eliminate this limitation with the use of a Fourier--Laplace transform. The 2D Fourier transform of Eq.~\eqref{eq:ESPRIT_t} is then expressed as
\begin{equation}
  R(\omega_1,\omega_2) = \sum_{i=1}^{N} {F} [\frac{a_i(t_2)}{i\omega_1 - b_i(t_2)}],
  \label{eq:ESPRIT_w}
\end{equation}
where ${F}$ represents the discrete Fourier transform (DFT) on $t_2$.  

Using  ESPRIT, we obtained the 2D IIR spectrum for the stretching--bending modes shown in Fig.~\ref{fgr:bending}. While conventional calculations require 192 sample points in the $t_1$ direction, we can reduce the  necessary calculations four times (48 sampling points) with the use of  ESPRIT, giving almost identical results. 

Note that while Prony's method has been incorporated into the HEOM formalism to eliminate instabilities arising from the discrete-bath HEOM\cite{dunn2019} and to apply the time-domain Prony fitting decomposition (t-PFD) scheme as an efficient description of SDF,\cite{chen2022}, the same method can also be used to evaluate a 2D spectral profile efficiently. However, we have found  ESPRIT to be more convenient and stable, in particular when the signal is mixed with noise.

\bibliography{TT22.bib,tanimura_publist.bib}

\begin{thebibliography}{83}%
\makeatletter
\providecommand \@ifxundefined [1]{%
 \@ifx{#1\undefined}
}%
\providecommand \@ifnum [1]{%
 \ifnum #1\expandafter \@firstoftwo
 \else \expandafter \@secondoftwo
 \fi
}%
\providecommand \@ifx [1]{%
 \ifx #1\expandafter \@firstoftwo
 \else \expandafter \@secondoftwo
 \fi
}%
\providecommand \natexlab [1]{#1}%
\providecommand \enquote  [1]{``#1''}%
\providecommand \bibnamefont  [1]{#1}%
\providecommand \bibfnamefont [1]{#1}%
\providecommand \citenamefont [1]{#1}%
\providecommand \href@noop [0]{\@secondoftwo}%
\providecommand \href [0]{\begingroup \@sanitize@url \@href}%
\providecommand \@href[1]{\@@startlink{#1}\@@href}%
\providecommand \@@href[1]{\endgroup#1\@@endlink}%
\providecommand \@sanitize@url [0]{\catcode `\\12\catcode `\$12\catcode
  `\&12\catcode `\#12\catcode `\^12\catcode `\_12\catcode `\%12\relax}%
\providecommand \@@startlink[1]{}%
\providecommand \@@endlink[0]{}%
\providecommand \url  [0]{\begingroup\@sanitize@url \@url }%
\providecommand \@url [1]{\endgroup\@href {#1}{\urlprefix }}%
\providecommand \urlprefix  [0]{URL }%
\providecommand \Eprint [0]{\href }%
\providecommand \doibase [0]{https://doi.org/}%
\providecommand \selectlanguage [0]{\@gobble}%
\providecommand \bibinfo  [0]{\@secondoftwo}%
\providecommand \bibfield  [0]{\@secondoftwo}%
\providecommand \translation [1]{[#1]}%
\providecommand \BibitemOpen [0]{}%
\providecommand \bibitemStop [0]{}%
\providecommand \bibitemNoStop [0]{.\EOS\space}%
\providecommand \EOS [0]{\spacefactor3000\relax}%
\providecommand \BibitemShut  [1]{\csname bibitem#1\endcsname}%
\let\auto@bib@innerbib\@empty
\bibitem [{\citenamefont {Ball}(1999)}]{Ball1999}%
  \BibitemOpen
  \bibfield  {author} {\bibinfo {author} {\bibfnamefont {P.}~\bibnamefont
  {Ball}},\ }\href@noop {} {\emph {\bibinfo {title} {Life's Matrix: A Biography
  of Water}}}\ (\bibinfo  {publisher} {Farrar, Straus and Giroux},\ \bibinfo
  {year} {1999})\BibitemShut {NoStop}%
\bibitem [{\citenamefont {Ohmine}\ and\ \citenamefont
  {Saito}(1999)}]{OCSACR1999}%
  \BibitemOpen
  \bibfield  {author} {\bibinfo {author} {\bibfnamefont {I.}~\bibnamefont
  {Ohmine}}\ and\ \bibinfo {author} {\bibfnamefont {S.}~\bibnamefont {Saito}},\
  }\bibfield  {title} {\enquote {\bibinfo {title} {Water dynamics {{:}}
  fluctuation, relaxation, and chemical reactions in hydrogen bond network
  rearrangement},}\ }\href {https://doi.org/10.1021/ar970161g} {\bibfield
  {journal} {\bibinfo  {journal} {Accounts of Chemical Research}\ }\textbf
  {\bibinfo {volume} {32}},\ \bibinfo {pages} {741--749} (\bibinfo {year}
  {1999})}\BibitemShut {NoStop}%
\bibitem [{\citenamefont {Bagchi}(2013)}]{bagchi_2013}%
  \BibitemOpen
  \bibfield  {author} {\bibinfo {author} {\bibfnamefont {B.}~\bibnamefont
  {Bagchi}},\ }\href {https://doi.org/10.1017/CBO9781139583947} {\emph
  {\bibinfo {title} {Water in Biological and Chemical Processes: From Structure
  and Dynamics to Function}}},\ Cambridge Molecular Science\ (\bibinfo
  {publisher} {Cambridge University Press},\ \bibinfo {year}
  {2013})\BibitemShut {NoStop}%
\bibitem [{\citenamefont {Fecko}\ \emph {et~al.}(2003)\citenamefont {Fecko},
  \citenamefont {Eaves}, \citenamefont {Loparo}, \citenamefont {Tokmakoff},\
  and\ \citenamefont {Geissler}}]{Tokmakoff2003H2O}%
  \BibitemOpen
  \bibfield  {author} {\bibinfo {author} {\bibfnamefont {C.}~\bibnamefont
  {Fecko}}, \bibinfo {author} {\bibfnamefont {J.}~\bibnamefont {Eaves}},
  \bibinfo {author} {\bibfnamefont {J.}~\bibnamefont {Loparo}}, \bibinfo
  {author} {\bibfnamefont {A.}~\bibnamefont {Tokmakoff}},\ and\ \bibinfo
  {author} {\bibfnamefont {P.}~\bibnamefont {Geissler}},\ }\bibfield  {title}
  {\enquote {\bibinfo {title} {Ultrafast hydrogen-bond dynamics in the infrared
  spectroscopy of water},}\ }\href {https://doi.org/10.1126/science.1087251}
  {\bibfield  {journal} {\bibinfo  {journal} {SCIENCE}\ }\textbf {\bibinfo
  {volume} {301}},\ \bibinfo {pages} {1698--1702} (\bibinfo {year}
  {2003})}\BibitemShut {NoStop}%
\bibitem [{\citenamefont {Rey}, \citenamefont {Møller},\ and\ \citenamefont
  {Hynes}(2004)}]{Hynes2004}%
  \BibitemOpen
  \bibfield  {author} {\bibinfo {author} {\bibfnamefont {R.}~\bibnamefont
  {Rey}}, \bibinfo {author} {\bibfnamefont {K.~B.}\ \bibnamefont {Møller}},\
  and\ \bibinfo {author} {\bibfnamefont {J.~T.}\ \bibnamefont {Hynes}},\
  }\bibfield  {title} {\enquote {\bibinfo {title} {Ultrafast vibrational
  population dynamics of water and related systems, a theoretical
  perspective},}\ }\href {https://doi.org/10.1021/cr020675f} {\bibfield
  {journal} {\bibinfo  {journal} {Chemical Reviews}\ }\textbf {\bibinfo
  {volume} {104}},\ \bibinfo {pages} {1915--1928} (\bibinfo {year}
  {2004})}\BibitemShut {NoStop}%
\bibitem [{\citenamefont {Ohmine}\ and\ \citenamefont
  {Tanaka}(1993)}]{Ohmine_ChemRev93}%
  \BibitemOpen
  \bibfield  {author} {\bibinfo {author} {\bibfnamefont {I.}~\bibnamefont
  {Ohmine}}\ and\ \bibinfo {author} {\bibfnamefont {H.}~\bibnamefont
  {Tanaka}},\ }\bibfield  {title} {\enquote {\bibinfo {title} {Fluctuation,
  relaxations, and hydration in liquid water. hydrogen-bond rearrangement
  dynamics},}\ }\href {https://doi.org/10.1021/cr00023a011} {\bibfield
  {journal} {\bibinfo  {journal} {Chemical Reviews}\ }\textbf {\bibinfo
  {volume} {93}},\ \bibinfo {pages} {2545--2566} (\bibinfo {year} {1993})},\
  \Eprint {https://arxiv.org/abs/https://doi.org/10.1021/cr00023a011}
  {https://doi.org/10.1021/cr00023a011} \BibitemShut {NoStop}%
\bibitem [{\citenamefont {Yagasaki}\ and\ \citenamefont
  {Saito}(2009)}]{Yagasaki_ACR42}%
  \BibitemOpen
  \bibfield  {author} {\bibinfo {author} {\bibfnamefont {T.}~\bibnamefont
  {Yagasaki}}\ and\ \bibinfo {author} {\bibfnamefont {S.}~\bibnamefont
  {Saito}},\ }\bibfield  {title} {\enquote {\bibinfo {title} {Molecular
  dynamics simulation of nonlinear spectroscopies of intermolecular motions in
  liquid water},}\ }\href {https://doi.org/10.1021/ar900007s} {\bibfield
  {journal} {\bibinfo  {journal} {Accounts of Chemical Research}\ }\textbf
  {\bibinfo {volume} {42}},\ \bibinfo {pages} {1250--1258} (\bibinfo {year}
  {2009})},\ \Eprint {https://arxiv.org/abs/https://doi.org/10.1021/ar900007s}
  {https://doi.org/10.1021/ar900007s} \BibitemShut {NoStop}%
\bibitem [{\citenamefont {Yagasaki}\ and\ \citenamefont
  {Saito}(2013)}]{Yagasaki_ARPC64}%
  \BibitemOpen
  \bibfield  {author} {\bibinfo {author} {\bibfnamefont {T.}~\bibnamefont
  {Yagasaki}}\ and\ \bibinfo {author} {\bibfnamefont {S.}~\bibnamefont
  {Saito}},\ }\bibfield  {title} {\enquote {\bibinfo {title} {Fluctuations and
  relaxation dynamics of liquid water revealed by linear and nonlinear
  spectroscopy},}\ }\href
  {https://doi.org/10.1146/annurev-physchem-040412-110150} {\bibfield
  {journal} {\bibinfo  {journal} {Annual Review of Physical Chemistry}\
  }\textbf {\bibinfo {volume} {64}},\ \bibinfo {pages} {55--75} (\bibinfo
  {year} {2013})},\ \Eprint
  {https://arxiv.org/abs/https://doi.org/10.1146/annurev-physchem-040412-110150}
  {https://doi.org/10.1146/annurev-physchem-040412-110150} \BibitemShut
  {NoStop}%
\bibitem [{\citenamefont {Ramasesha}\ \emph {et~al.}(2013)\citenamefont
  {Ramasesha}, \citenamefont {De~Marco}, \citenamefont {Mandal},\ and\
  \citenamefont {Tokmakoff}}]{TokmakoffNat2013}%
  \BibitemOpen
  \bibfield  {author} {\bibinfo {author} {\bibfnamefont {K.}~\bibnamefont
  {Ramasesha}}, \bibinfo {author} {\bibfnamefont {L.}~\bibnamefont {De~Marco}},
  \bibinfo {author} {\bibfnamefont {A.}~\bibnamefont {Mandal}},\ and\ \bibinfo
  {author} {\bibfnamefont {A.}~\bibnamefont {Tokmakoff}},\ }\bibfield  {title}
  {\enquote {\bibinfo {title} {Water vibrations have strongly mixed intra- and
  intermolecular character},}\ }\href {https://doi.org/10.1038/nchem.1757}
  {\bibfield  {journal} {\bibinfo  {journal} {Nature Chemistry}\ }\textbf
  {\bibinfo {volume} {5}},\ \bibinfo {pages} {935--940} (\bibinfo {year}
  {2013})}\BibitemShut {NoStop}%
\bibitem [{\citenamefont {Bertie}\ and\ \citenamefont
  {Lan}(1996)}]{Bertie96IRexp}%
  \BibitemOpen
  \bibfield  {author} {\bibinfo {author} {\bibfnamefont {J.~E.}\ \bibnamefont
  {Bertie}}\ and\ \bibinfo {author} {\bibfnamefont {Z.}~\bibnamefont {Lan}},\
  }\bibfield  {title} {\enquote {\bibinfo {title} {Infrared intensities of
  liquids xx: The intensity of the oh stretching band of liquid water
  revisited, and the best current values of the optical constants of h2o(l) at
  25{\textdegree}c between 15,000 and 1 cm-1},}\ }\href
  {https://opg.optica.org/as/abstract.cfm?URI=as-50-8-1047} {\bibfield
  {journal} {\bibinfo  {journal} {Appl. Spectrosc.}\ }\textbf {\bibinfo
  {volume} {50}},\ \bibinfo {pages} {1047--1057} (\bibinfo {year}
  {1996})}\BibitemShut {NoStop}%
\bibitem [{\citenamefont {Maréchal}(2011)}]{IRexp2011}%
  \BibitemOpen
  \bibfield  {author} {\bibinfo {author} {\bibfnamefont {Y.}~\bibnamefont
  {Maréchal}},\ }\bibfield  {title} {\enquote {\bibinfo {title} {The molecular
  structure of liquid water delivered by absorption spectroscopy in the whole
  ir region completed with thermodynamics data},}\ }\href
  {https://doi.org/https://doi.org/10.1016/j.molstruc.2011.07.054} {\bibfield
  {journal} {\bibinfo  {journal} {Journal of Molecular Structure}\ }\textbf
  {\bibinfo {volume} {1004}},\ \bibinfo {pages} {146--155} (\bibinfo {year}
  {2011})}\BibitemShut {NoStop}%
\bibitem [{\citenamefont {Brooker}\ \emph {et~al.}(1989)\citenamefont
  {Brooker}, \citenamefont {Hancock}, \citenamefont {Rice},\ and\ \citenamefont
  {Shapter}}]{Brooker1989Ramanexp}%
  \BibitemOpen
  \bibfield  {author} {\bibinfo {author} {\bibfnamefont {M.~H.}\ \bibnamefont
  {Brooker}}, \bibinfo {author} {\bibfnamefont {G.~W.}\ \bibnamefont
  {Hancock}}, \bibinfo {author} {\bibfnamefont {B.}~\bibnamefont {Rice}},\ and\
  \bibinfo {author} {\bibfnamefont {J.~G.}\ \bibnamefont {Shapter}},\
  }\bibfield  {title} {\enquote {\bibinfo {title} {\uppercase{R}aman frequency
  and intensity studies of liquid h2o, h218o and d2o},}\ }\href@noop {}
  {\bibfield  {journal} {\bibinfo  {journal} {Journal of Raman Spectroscopy}\
  }\textbf {\bibinfo {volume} {20}},\ \bibinfo {pages} {683--694} (\bibinfo
  {year} {1989})}\BibitemShut {NoStop}%
\bibitem [{\citenamefont {Pattenaude}, \citenamefont {Streacker},\ and\
  \citenamefont {Ben-Amotz}(2018)}]{Raman2018exp}%
  \BibitemOpen
  \bibfield  {author} {\bibinfo {author} {\bibfnamefont {S.~R.}\ \bibnamefont
  {Pattenaude}}, \bibinfo {author} {\bibfnamefont {L.~M.}\ \bibnamefont
  {Streacker}},\ and\ \bibinfo {author} {\bibfnamefont {D.}~\bibnamefont
  {Ben-Amotz}},\ }\bibfield  {title} {\enquote {\bibinfo {title} {Temperature
  and polarization dependent \uppercase{R}aman spectra of liquid h2o and
  d2o},}\ }\href {https://doi.org/https://doi.org/10.1002/jrs.5465} {\bibfield
  {journal} {\bibinfo  {journal} {Journal of Raman Spectroscopy}\ }\textbf
  {\bibinfo {volume} {49}},\ \bibinfo {pages} {1860--1866} (\bibinfo {year}
  {2018})}\BibitemShut {NoStop}%
\bibitem [{\citenamefont {Tanimura}\ and\ \citenamefont
  {Mukamel}(1993)}]{TM93JCP}%
  \BibitemOpen
  \bibfield  {author} {\bibinfo {author} {\bibfnamefont {Y.}~\bibnamefont
  {Tanimura}}\ and\ \bibinfo {author} {\bibfnamefont {S.}~\bibnamefont
  {Mukamel}},\ }\bibfield  {title} {\enquote {\bibinfo {title} {Two-dimensional
  femtosecond vibrational spectroscopy of liquids},}\ }\href
  {https://doi.org/10.1063/1.465484} {\bibfield  {journal} {\bibinfo  {journal}
  {The Journal of Chemical Physics}\ }\textbf {\bibinfo {volume} {99}},\
  \bibinfo {pages} {9496--9511} (\bibinfo {year} {1993})}\BibitemShut {NoStop}%
\bibitem [{\citenamefont {Mukamel}(1999)}]{mukamel1999principles}%
  \BibitemOpen
  \bibfield  {author} {\bibinfo {author} {\bibfnamefont {S.}~\bibnamefont
  {Mukamel}},\ }\href@noop {} {\emph {\bibinfo {title} {Principles of nonlinear
  optical spectroscopy}}},\ \bibinfo {number} {6}\ (\bibinfo  {publisher}
  {Oxford University Press on Demand},\ \bibinfo {year} {1999})\BibitemShut
  {NoStop}%
\bibitem [{\citenamefont {Cho}(2009)}]{Cho2009}%
  \BibitemOpen
  \bibfield  {author} {\bibinfo {author} {\bibfnamefont {M.}~\bibnamefont
  {Cho}},\ }\href {https://doi.org/https://doi.org/10.1201/9781420084306}
  {\emph {\bibinfo {title} {Two-Dimensional Optical Spectroscopy}}}\ (\bibinfo
  {publisher} {CRC Press},\ \bibinfo {year} {2009})\BibitemShut {NoStop}%
\bibitem [{\citenamefont {Hamm}\ and\ \citenamefont
  {Zanni}(2011)}]{Hamm2011ConceptsAM}%
  \BibitemOpen
  \bibfield  {author} {\bibinfo {author} {\bibfnamefont {P.}~\bibnamefont
  {Hamm}}\ and\ \bibinfo {author} {\bibfnamefont {M.~T.}\ \bibnamefont
  {Zanni}},\ }\href {https://doi.org/https://doi.org/10.1017/CBO9780511675935}
  {\emph {\bibinfo {title} {Concepts and Methods of 2D Infrared
  Spectroscopy}}}\ (\bibinfo  {publisher} {Cambridge University Press},\
  \bibinfo {year} {2011})\BibitemShut {NoStop}%
\bibitem [{\citenamefont {De~Marco}\ \emph {et~al.}(2016)\citenamefont
  {De~Marco}, \citenamefont {Fournier}, \citenamefont {Thämer}, \citenamefont
  {Carpenter},\ and\ \citenamefont {Tokmakoff}}]{Tokmakoff2016H2O}%
  \BibitemOpen
  \bibfield  {author} {\bibinfo {author} {\bibfnamefont {L.}~\bibnamefont
  {De~Marco}}, \bibinfo {author} {\bibfnamefont {J.~A.}\ \bibnamefont
  {Fournier}}, \bibinfo {author} {\bibfnamefont {M.}~\bibnamefont {Thämer}},
  \bibinfo {author} {\bibfnamefont {W.}~\bibnamefont {Carpenter}},\ and\
  \bibinfo {author} {\bibfnamefont {A.}~\bibnamefont {Tokmakoff}},\ }\bibfield
  {title} {\enquote {\bibinfo {title} {Anharmonic exciton dynamics and energy
  dissipation in liquid water from two-dimensional infrared spectroscopy},}\
  }\href {https://doi.org/10.1063/1.4961752} {\bibfield  {journal} {\bibinfo
  {journal} {The Journal of Chemical Physics}\ }\textbf {\bibinfo {volume}
  {145}},\ \bibinfo {pages} {094501} (\bibinfo {year} {2016})},\ \Eprint
  {https://arxiv.org/abs/https://aip.scitation.org/doi/pdf/10.1063/1.4961752}
  {https://aip.scitation.org/doi/pdf/10.1063/1.4961752} \BibitemShut {NoStop}%
\bibitem [{\citenamefont {Lewis}\ \emph {et~al.}(2022)\citenamefont {Lewis},
  \citenamefont {Dereka}, \citenamefont {Zhang}, \citenamefont {Maginn},\ and\
  \citenamefont {Tokmakoff}}]{Tokmakoff2022}%
  \BibitemOpen
  \bibfield  {author} {\bibinfo {author} {\bibfnamefont {N.~H.~C.}\
  \bibnamefont {Lewis}}, \bibinfo {author} {\bibfnamefont {B.}~\bibnamefont
  {Dereka}}, \bibinfo {author} {\bibfnamefont {Y.}~\bibnamefont {Zhang}},
  \bibinfo {author} {\bibfnamefont {E.~J.}\ \bibnamefont {Maginn}},\ and\
  \bibinfo {author} {\bibfnamefont {A.}~\bibnamefont {Tokmakoff}},\ }\bibfield
  {title} {\enquote {\bibinfo {title} {From networked to isolated: Observing
  water hydrogen bonds in concentrated electrolytes with two-dimensional
  infrared spectroscopy},}\ }\href {https://doi.org/10.1021/acs.jpcb.2c03341}
  {\bibfield  {journal} {\bibinfo  {journal} {The Journal of Physical Chemistry
  B}\ }\textbf {\bibinfo {volume} {126}},\ \bibinfo {pages} {5305--5319}
  (\bibinfo {year} {2022})},\ \bibinfo {note} {pMID: 35829623},\ \Eprint
  {https://arxiv.org/abs/https://doi.org/10.1021/acs.jpcb.2c03341}
  {https://doi.org/10.1021/acs.jpcb.2c03341} \BibitemShut {NoStop}%
\bibitem [{\citenamefont {Huse}\ \emph {et~al.}(2005)\citenamefont {Huse},
  \citenamefont {Ashihara}, \citenamefont {Nibbering},\ and\ \citenamefont
  {Elsaesser}}]{ElsaesserCPL2005}%
  \BibitemOpen
  \bibfield  {author} {\bibinfo {author} {\bibfnamefont {N.}~\bibnamefont
  {Huse}}, \bibinfo {author} {\bibfnamefont {S.}~\bibnamefont {Ashihara}},
  \bibinfo {author} {\bibfnamefont {E.}~\bibnamefont {Nibbering}},\ and\
  \bibinfo {author} {\bibfnamefont {T.}~\bibnamefont {Elsaesser}},\ }\bibfield
  {title} {\enquote {\bibinfo {title} {Ultrafast vibrational relaxation of o-h
  bending and librational excitations in liquid h2o},}\ }\href
  {https://doi.org/10.1016/j.cplett.2005.02.007} {\bibfield  {journal}
  {\bibinfo  {journal} {CHEMICAL PHYSICS LETTERS}\ }\textbf {\bibinfo {volume}
  {404}},\ \bibinfo {pages} {389--393} (\bibinfo {year} {2005})}\BibitemShut
  {NoStop}%
\bibitem [{\citenamefont {Cowan}\ \emph {et~al.}(2005)\citenamefont {Cowan},
  \citenamefont {Bruner}, \citenamefont {Huse}, \citenamefont {Dwyer},
  \citenamefont {Chugh}, \citenamefont {Nibbering}, \citenamefont {Elsaesser},\
  and\ \citenamefont {Miller}}]{ElsaesserH2O}%
  \BibitemOpen
  \bibfield  {author} {\bibinfo {author} {\bibfnamefont {M.}~\bibnamefont
  {Cowan}}, \bibinfo {author} {\bibfnamefont {B.}~\bibnamefont {Bruner}},
  \bibinfo {author} {\bibfnamefont {N.}~\bibnamefont {Huse}}, \bibinfo {author}
  {\bibfnamefont {J.}~\bibnamefont {Dwyer}}, \bibinfo {author} {\bibfnamefont
  {B.}~\bibnamefont {Chugh}}, \bibinfo {author} {\bibfnamefont
  {E.}~\bibnamefont {Nibbering}}, \bibinfo {author} {\bibfnamefont
  {T.}~\bibnamefont {Elsaesser}},\ and\ \bibinfo {author} {\bibfnamefont
  {R.}~\bibnamefont {Miller}},\ }\bibfield  {title} {\enquote {\bibinfo {title}
  {Ultrafast memory loss and energy redistribution in the hydrogen bond network
  of liquid h2o},}\ }\href {https://doi.org/10.1038/nature03383} {\bibfield
  {journal} {\bibinfo  {journal} {NATURE}\ }\textbf {\bibinfo {volume} {434}},\
  \bibinfo {pages} {199--202} (\bibinfo {year} {2005})}\BibitemShut {NoStop}%
\bibitem [{\citenamefont {Ashihara}\ \emph {et~al.}(2007)\citenamefont
  {Ashihara}, \citenamefont {Huse}, \citenamefont {Espagne}, \citenamefont
  {Nibbering},\ and\ \citenamefont {Elsaesser}}]{ElsaesserJPCA2007}%
  \BibitemOpen
  \bibfield  {author} {\bibinfo {author} {\bibfnamefont {S.}~\bibnamefont
  {Ashihara}}, \bibinfo {author} {\bibfnamefont {N.}~\bibnamefont {Huse}},
  \bibinfo {author} {\bibfnamefont {A.}~\bibnamefont {Espagne}}, \bibinfo
  {author} {\bibfnamefont {E.~T.~J.}\ \bibnamefont {Nibbering}},\ and\ \bibinfo
  {author} {\bibfnamefont {T.}~\bibnamefont {Elsaesser}},\ }\bibfield  {title}
  {\enquote {\bibinfo {title} {Ultrafast structural dynamics of water induced
  by dissipation of vibrational energy},}\ }\href
  {https://doi.org/10.1021/jp0676538} {\bibfield  {journal} {\bibinfo
  {journal} {JOURNAL OF PHYSICAL CHEMISTRY A}\ }\textbf {\bibinfo {volume}
  {111}},\ \bibinfo {pages} {743--746} (\bibinfo {year} {2007})}\BibitemShut
  {NoStop}%
\bibitem [{\citenamefont {Hamm}\ and\ \citenamefont
  {Savolainen}(2012)}]{HammTHz2012}%
  \BibitemOpen
  \bibfield  {author} {\bibinfo {author} {\bibfnamefont {P.}~\bibnamefont
  {Hamm}}\ and\ \bibinfo {author} {\bibfnamefont {J.}~\bibnamefont
  {Savolainen}},\ }\bibfield  {title} {\enquote {\bibinfo {title}
  {Two-dimensional-\uppercase{R}aman-\uppercase{T}erahertz spectroscopy of
  water: Theory},}\ }\href {https://doi.org/10.1063/1.3691601} {\bibfield
  {journal} {\bibinfo  {journal} {The Journal of Chemical Physics}\ }\textbf
  {\bibinfo {volume} {136}},\ \bibinfo {pages} {094516} (\bibinfo {year}
  {2012})},\ \Eprint {https://arxiv.org/abs/https://doi.org/10.1063/1.3691601}
  {https://doi.org/10.1063/1.3691601} \BibitemShut {NoStop}%
\bibitem [{\citenamefont {Savolainen}, \citenamefont {Ahmed},\ and\
  \citenamefont {Hamm}(2013)}]{Hamm2013PNAS}%
  \BibitemOpen
  \bibfield  {author} {\bibinfo {author} {\bibfnamefont {J.}~\bibnamefont
  {Savolainen}}, \bibinfo {author} {\bibfnamefont {S.}~\bibnamefont {Ahmed}},\
  and\ \bibinfo {author} {\bibfnamefont {P.}~\bibnamefont {Hamm}},\ }\bibfield
  {title} {\enquote {\bibinfo {title} {Two-dimensional
  \uppercase{R}aman-\uppercase{T}erahertz spectroscopy of water},}\ }\href
  {https://doi.org/10.1073/pnas.1317459110} {\bibfield  {journal} {\bibinfo
  {journal} {Proceedings of the National Academy of Sciences}\ }\textbf
  {\bibinfo {volume} {110}},\ \bibinfo {pages} {20402--20407} (\bibinfo {year}
  {2013})},\ \Eprint
  {https://arxiv.org/abs/https://www.pnas.org/doi/pdf/10.1073/pnas.1317459110}
  {https://www.pnas.org/doi/pdf/10.1073/pnas.1317459110} \BibitemShut {NoStop}%
\bibitem [{\citenamefont {Hamm}\ and\ \citenamefont
  {Shalit}(2017)}]{HammPerspH2O2017}%
  \BibitemOpen
  \bibfield  {author} {\bibinfo {author} {\bibfnamefont {P.}~\bibnamefont
  {Hamm}}\ and\ \bibinfo {author} {\bibfnamefont {A.}~\bibnamefont {Shalit}},\
  }\bibfield  {title} {\enquote {\bibinfo {title} {Perspective: Echoes in
  2d-\uppercase{R}aman-\uppercase{TH}z spectroscopy},}\ }\href
  {https://doi.org/10.1063/1.4979288} {\bibfield  {journal} {\bibinfo
  {journal} {The Journal of Chemical Physics}\ }\textbf {\bibinfo {volume}
  {146}},\ \bibinfo {pages} {130901} (\bibinfo {year} {2017})},\ \Eprint
  {https://arxiv.org/abs/https://doi.org/10.1063/1.4979288}
  {https://doi.org/10.1063/1.4979288} \BibitemShut {NoStop}%
\bibitem [{\citenamefont {Grechko}\ \emph {et~al.}(2018)\citenamefont
  {Grechko}, \citenamefont {Hasegawa}, \citenamefont {D'Angelo}, \citenamefont
  {Ito}, \citenamefont {Turchinovich}, \citenamefont {Nagata},\ and\
  \citenamefont {Bonn}}]{grechko2018}%
  \BibitemOpen
  \bibfield  {author} {\bibinfo {author} {\bibfnamefont {M.}~\bibnamefont
  {Grechko}}, \bibinfo {author} {\bibfnamefont {T.}~\bibnamefont {Hasegawa}},
  \bibinfo {author} {\bibfnamefont {F.}~\bibnamefont {D'Angelo}}, \bibinfo
  {author} {\bibfnamefont {H.}~\bibnamefont {Ito}}, \bibinfo {author}
  {\bibfnamefont {D.}~\bibnamefont {Turchinovich}}, \bibinfo {author}
  {\bibfnamefont {Y.}~\bibnamefont {Nagata}},\ and\ \bibinfo {author}
  {\bibfnamefont {M.}~\bibnamefont {Bonn}},\ }\bibfield  {title} {\enquote
  {\bibinfo {title} {Coupling between intra- and intermolecular motions in
  liquid water revealed by two-dimensional
  \uppercase{T}erahertz-infrared-visible spectroscopy},}\ }\href
  {https://doi.org/10.1038/s41467-018-03303-y} {\bibfield  {journal} {\bibinfo
  {journal} {Nat Commun}\ }\textbf {\bibinfo {volume} {9}},\ \bibinfo {pages}
  {885} (\bibinfo {year} {2018})}\BibitemShut {NoStop}%
\bibitem [{\citenamefont {Vietze}\ \emph {et~al.}(2021)\citenamefont {Vietze},
  \citenamefont {Backus}, \citenamefont {Bonn},\ and\ \citenamefont
  {Grechko}}]{Bonn2DTZIFvis2021}%
  \BibitemOpen
  \bibfield  {author} {\bibinfo {author} {\bibfnamefont {L.}~\bibnamefont
  {Vietze}}, \bibinfo {author} {\bibfnamefont {E.~H.~G.}\ \bibnamefont
  {Backus}}, \bibinfo {author} {\bibfnamefont {M.}~\bibnamefont {Bonn}},\ and\
  \bibinfo {author} {\bibfnamefont {M.}~\bibnamefont {Grechko}},\ }\bibfield
  {title} {\enquote {\bibinfo {title} {Distinguishing different excitation
  pathways in two-dimensional \uppercase{T}erahertz-infrared-visible
  spectroscopy},}\ }\href {https://doi.org/10.1063/5.0047918} {\bibfield
  {journal} {\bibinfo  {journal} {The Journal of Chemical Physics}\ }\textbf
  {\bibinfo {volume} {154}},\ \bibinfo {pages} {174201} (\bibinfo {year}
  {2021})},\ \Eprint {https://arxiv.org/abs/https://doi.org/10.1063/5.0047918}
  {https://doi.org/10.1063/5.0047918} \BibitemShut {NoStop}%
\bibitem [{\citenamefont {Jansen}\ \emph {et~al.}(2019)\citenamefont {Jansen},
  \citenamefont {Saito}, \citenamefont {Jeon},\ and\ \citenamefont
  {Cho}}]{JansenChoShinji2DVPerspe2019}%
  \BibitemOpen
  \bibfield  {author} {\bibinfo {author} {\bibfnamefont {T.~l.~C.}\
  \bibnamefont {Jansen}}, \bibinfo {author} {\bibfnamefont {S.}~\bibnamefont
  {Saito}}, \bibinfo {author} {\bibfnamefont {J.}~\bibnamefont {Jeon}},\ and\
  \bibinfo {author} {\bibfnamefont {M.}~\bibnamefont {Cho}},\ }\bibfield
  {title} {\enquote {\bibinfo {title} {Theory of coherent two-dimensional
  vibrational spectroscopy},}\ }\href {https://doi.org/10.1063/1.5083966}
  {\bibfield  {journal} {\bibinfo  {journal} {The Journal of Chemical Physics}\
  }\textbf {\bibinfo {volume} {150}},\ \bibinfo {pages} {100901} (\bibinfo
  {year} {2019})},\ \Eprint
  {https://arxiv.org/abs/https://doi.org/10.1063/1.5083966}
  {https://doi.org/10.1063/1.5083966} \BibitemShut {NoStop}%
\bibitem [{\citenamefont {Hasegawa}\ and\ \citenamefont
  {Tanimura}(2011)}]{HT11JPCB}%
  \BibitemOpen
  \bibfield  {author} {\bibinfo {author} {\bibfnamefont {T.}~\bibnamefont
  {Hasegawa}}\ and\ \bibinfo {author} {\bibfnamefont {Y.}~\bibnamefont
  {Tanimura}},\ }\bibfield  {title} {\enquote {\bibinfo {title} {A polarizable
  water model for intramolecular and intermolecular vibrational
  spectroscopies},}\ }\href {https://doi.org/10.1021/jp111308f} {\bibfield
  {journal} {\bibinfo  {journal} {The Journal of Physical Chemistry B}\
  }\textbf {\bibinfo {volume} {115}},\ \bibinfo {pages} {5545--5553} (\bibinfo
  {year} {2011})}\BibitemShut {NoStop}%
\bibitem [{\citenamefont {Imoto}, \citenamefont {Xantheas},\ and\ \citenamefont
  {Saito}(2013)}]{ImotXanteasSaitoJCP2013H2O}%
  \BibitemOpen
  \bibfield  {author} {\bibinfo {author} {\bibfnamefont {S.}~\bibnamefont
  {Imoto}}, \bibinfo {author} {\bibfnamefont {S.~S.}\ \bibnamefont
  {Xantheas}},\ and\ \bibinfo {author} {\bibfnamefont {S.}~\bibnamefont
  {Saito}},\ }\bibfield  {title} {\enquote {\bibinfo {title} {Ultrafast
  dynamics of liquid water: Frequency fluctuations of the oh stretch and the
  hoh bend},}\ }\href {https://doi.org/10.1063/1.4813071} {\bibfield  {journal}
  {\bibinfo  {journal} {The Journal of Chemical Physics}\ }\textbf {\bibinfo
  {volume} {139}},\ \bibinfo {pages} {044503} (\bibinfo {year} {2013})},\
  \Eprint {https://arxiv.org/abs/https://doi.org/10.1063/1.4813071}
  {https://doi.org/10.1063/1.4813071} \BibitemShut {NoStop}%
\bibitem [{\citenamefont {Jeon}\ and\ \citenamefont
  {Cho}(2014)}]{ChoH2OMD2014}%
  \BibitemOpen
  \bibfield  {author} {\bibinfo {author} {\bibfnamefont {J.}~\bibnamefont
  {Jeon}}\ and\ \bibinfo {author} {\bibfnamefont {M.}~\bibnamefont {Cho}},\
  }\bibfield  {title} {\enquote {\bibinfo {title} {An accurate classical
  simulation of a two-dimensional vibrational spectrum: Od stretch spectrum of
  a hydrated hod molecule},}\ }\href {https://doi.org/10.1021/jp501182d}
  {\bibfield  {journal} {\bibinfo  {journal} {The Journal of Physical Chemistry
  B}\ }\textbf {\bibinfo {volume} {118}},\ \bibinfo {pages} {8148--8161}
  (\bibinfo {year} {2014})},\ \bibinfo {note} {pMID: 24601590},\ \Eprint
  {https://arxiv.org/abs/https://doi.org/10.1021/jp501182d}
  {https://doi.org/10.1021/jp501182d} \BibitemShut {NoStop}%
\bibitem [{\citenamefont {Medders}, \citenamefont {Babin},\ and\ \citenamefont
  {Paesani}(2014)}]{PaesaniJCTC2014H2O}%
  \BibitemOpen
  \bibfield  {author} {\bibinfo {author} {\bibfnamefont {G.~R.}\ \bibnamefont
  {Medders}}, \bibinfo {author} {\bibfnamefont {V.}~\bibnamefont {Babin}},\
  and\ \bibinfo {author} {\bibfnamefont {F.}~\bibnamefont {Paesani}},\
  }\bibfield  {title} {\enquote {\bibinfo {title} {Development of a
  “first-principles” water potential with flexible monomers. iii. liquid
  phase properties},}\ }\href {https://doi.org/10.1021/ct5004115} {\bibfield
  {journal} {\bibinfo  {journal} {Journal of Chemical Theory and Computation}\
  }\textbf {\bibinfo {volume} {10}},\ \bibinfo {pages} {2906--2910} (\bibinfo
  {year} {2014})},\ \bibinfo {note} {pMID: 26588266},\ \Eprint
  {https://arxiv.org/abs/https://doi.org/10.1021/ct5004115}
  {https://doi.org/10.1021/ct5004115} \BibitemShut {NoStop}%
\bibitem [{\citenamefont {Ishizaki}\ and\ \citenamefont
  {Tanimura}(2006)}]{IT06JCP}%
  \BibitemOpen
  \bibfield  {author} {\bibinfo {author} {\bibfnamefont {A.}~\bibnamefont
  {Ishizaki}}\ and\ \bibinfo {author} {\bibfnamefont {Y.}~\bibnamefont
  {Tanimura}},\ }\bibfield  {title} {\enquote {\bibinfo {title} {Modeling
  vibrational dephasing and energy relaxation of intramolecular anharmonic
  modes for multidimensional infrared spectroscopies},}\ }\href
  {https://doi.org/10.1063/1.2244558} {\bibfield  {journal} {\bibinfo
  {journal} {The Journal of Chemical Physics}\ }\textbf {\bibinfo {volume}
  {125}},\ \bibinfo {pages} {084501} (\bibinfo {year} {2006})}\BibitemShut
  {NoStop}%
\bibitem [{\citenamefont {Sakurai}\ and\ \citenamefont
  {Tanimura}(2011)}]{ST11JPCA}%
  \BibitemOpen
  \bibfield  {author} {\bibinfo {author} {\bibfnamefont {A.}~\bibnamefont
  {Sakurai}}\ and\ \bibinfo {author} {\bibfnamefont {Y.}~\bibnamefont
  {Tanimura}},\ }\bibfield  {title} {\enquote {\bibinfo {title} {Does $\hbar$
  play a role in multidimensional spectroscopy? reduced hierarchy equations of
  motion approach to molecular vibrations},}\ }\href
  {https://doi.org/10.1021/jp1095618} {\bibfield  {journal} {\bibinfo
  {journal} {The Journal of Physical Chemistry A}\ }\textbf {\bibinfo {volume}
  {115}},\ \bibinfo {pages} {4009--4022} (\bibinfo {year} {2011})}\BibitemShut
  {NoStop}%
\bibitem [{\citenamefont {Liu}\ \emph {et~al.}(2011)\citenamefont {Liu},
  \citenamefont {Miller}, \citenamefont {Fanourgakis}, \citenamefont
  {Xantheas}, \citenamefont {Imoto},\ and\ \citenamefont
  {Saito}}]{Imoto_JCP135}%
  \BibitemOpen
  \bibfield  {author} {\bibinfo {author} {\bibfnamefont {J.}~\bibnamefont
  {Liu}}, \bibinfo {author} {\bibfnamefont {W.~H.}\ \bibnamefont {Miller}},
  \bibinfo {author} {\bibfnamefont {G.~S.}\ \bibnamefont {Fanourgakis}},
  \bibinfo {author} {\bibfnamefont {S.~S.}\ \bibnamefont {Xantheas}}, \bibinfo
  {author} {\bibfnamefont {S.}~\bibnamefont {Imoto}},\ and\ \bibinfo {author}
  {\bibfnamefont {S.}~\bibnamefont {Saito}},\ }\bibfield  {title} {\enquote
  {\bibinfo {title} {Insights in quantum dynamical effects in the infrared
  spectroscopy of liquid water from a semiclassical study with an ab
  initio-based flexible and polarizable force field},}\ }\href
  {https://doi.org/10.1063/1.3670960} {\bibfield  {journal} {\bibinfo
  {journal} {The Journal of Chemical Physics}\ }\textbf {\bibinfo {volume}
  {135}},\ \bibinfo {pages} {244503} (\bibinfo {year} {2011})},\ \Eprint
  {https://arxiv.org/abs/https://doi.org/10.1063/1.3670960}
  {https://doi.org/10.1063/1.3670960} \BibitemShut {NoStop}%
\bibitem [{\citenamefont {Rossi}\ \emph {et~al.}(2014)\citenamefont {Rossi},
  \citenamefont {Liu}, \citenamefont {Paesani}, \citenamefont {Bowman},\ and\
  \citenamefont {Ceriotti}}]{PaesaniJCP2014qMD}%
  \BibitemOpen
  \bibfield  {author} {\bibinfo {author} {\bibfnamefont {M.}~\bibnamefont
  {Rossi}}, \bibinfo {author} {\bibfnamefont {H.}~\bibnamefont {Liu}}, \bibinfo
  {author} {\bibfnamefont {F.}~\bibnamefont {Paesani}}, \bibinfo {author}
  {\bibfnamefont {J.}~\bibnamefont {Bowman}},\ and\ \bibinfo {author}
  {\bibfnamefont {M.}~\bibnamefont {Ceriotti}},\ }\bibfield  {title} {\enquote
  {\bibinfo {title} {Communication: On the consistency of approximate quantum
  dynamics simulation methods for vibrational spectra in the condensed
  phase},}\ }\href {https://doi.org/10.1063/1.4901214} {\bibfield  {journal}
  {\bibinfo  {journal} {The Journal of Chemical Physics}\ }\textbf {\bibinfo
  {volume} {141}},\ \bibinfo {pages} {181101} (\bibinfo {year} {2014})},\
  \Eprint {https://arxiv.org/abs/https://doi.org/10.1063/1.4901214}
  {https://doi.org/10.1063/1.4901214} \BibitemShut {NoStop}%
\bibitem [{\citenamefont {Saito}\ and\ \citenamefont
  {Ohmine}(2006)}]{Shinji2DRaman2006}%
  \BibitemOpen
  \bibfield  {author} {\bibinfo {author} {\bibfnamefont {S.}~\bibnamefont
  {Saito}}\ and\ \bibinfo {author} {\bibfnamefont {I.}~\bibnamefont {Ohmine}},\
  }\bibfield  {title} {\enquote {\bibinfo {title} {Fifth-order two-dimensional
  \uppercase{R}aman spectroscopy of liquid water, crystalline ice ih and
  amorphous ices: Sensitivity to anharmonic dynamics and local hydrogen bond
  network structure},}\ }\href {https://doi.org/10.1063/1.2232254} {\bibfield
  {journal} {\bibinfo  {journal} {The Journal of Chemical Physics}\ }\textbf
  {\bibinfo {volume} {125}},\ \bibinfo {pages} {084506} (\bibinfo {year}
  {2006})},\ \Eprint {https://arxiv.org/abs/https://doi.org/10.1063/1.2232254}
  {https://doi.org/10.1063/1.2232254} \BibitemShut {NoStop}%
\bibitem [{\citenamefont {Hamm}(2014)}]{hamm2014}%
  \BibitemOpen
  \bibfield  {author} {\bibinfo {author} {\bibfnamefont {P.}~\bibnamefont
  {Hamm}},\ }\bibfield  {title} {\enquote {\bibinfo {title}
  {{{2D-\uppercase{R}aman-\uppercase{TH}z}} spectroscopy: {{A}} sensitive test
  of polarizable water models},}\ }\href {https://doi.org/10.1063/1.4901216}
  {\bibfield  {journal} {\bibinfo  {journal} {The Journal of Chemical Physics}\
  }\textbf {\bibinfo {volume} {141}},\ \bibinfo {pages} {184201} (\bibinfo
  {year} {2014})}\BibitemShut {NoStop}%
\bibitem [{\citenamefont {Ito}, \citenamefont {Hasegawa},\ and\ \citenamefont
  {Tanimura}(2014)}]{IHT14JCP}%
  \BibitemOpen
  \bibfield  {author} {\bibinfo {author} {\bibfnamefont {H.}~\bibnamefont
  {Ito}}, \bibinfo {author} {\bibfnamefont {T.}~\bibnamefont {Hasegawa}},\ and\
  \bibinfo {author} {\bibfnamefont {Y.}~\bibnamefont {Tanimura}},\ }\bibfield
  {title} {\enquote {\bibinfo {title} {Calculating two-dimensional
  \uppercase{TH}z-\uppercase{R}aman-\uppercase{TH}z and
  \uppercase{R}aman-\uppercase{TH}z-\uppercase{TH}z signals for various
  molecular liquids: The samplers},}\ }\href
  {https://doi.org/10.1063/1.4895908} {\bibfield  {journal} {\bibinfo
  {journal} {The Journal of Chemical Physics}\ }\textbf {\bibinfo {volume}
  {141}},\ \bibinfo {pages} {124503} (\bibinfo {year} {2014})}\BibitemShut
  {NoStop}%
\bibitem [{\citenamefont {Ito}, \citenamefont {Hasegawa},\ and\ \citenamefont
  {Tanimura}(2016)}]{IHT16JPCL}%
  \BibitemOpen
  \bibfield  {author} {\bibinfo {author} {\bibfnamefont {H.}~\bibnamefont
  {Ito}}, \bibinfo {author} {\bibfnamefont {T.}~\bibnamefont {Hasegawa}},\ and\
  \bibinfo {author} {\bibfnamefont {Y.}~\bibnamefont {Tanimura}},\ }\bibfield
  {title} {\enquote {\bibinfo {title} {Effects of intermolecular charge
  transfer in liquid water on \uppercase{R}aman spectra},}\ }\href
  {https://doi.org/10.1021/acs.jpclett.6b01766} {\bibfield  {journal} {\bibinfo
   {journal} {The Journal of Physical Chemistry Letters}\ }\textbf {\bibinfo
  {volume} {7}},\ \bibinfo {pages} {4147--4151} (\bibinfo {year}
  {2016})}\BibitemShut {NoStop}%
\bibitem [{\citenamefont {Hasegawa}\ and\ \citenamefont
  {Tanimura}(2008)}]{HT08JCP}%
  \BibitemOpen
  \bibfield  {author} {\bibinfo {author} {\bibfnamefont {T.}~\bibnamefont
  {Hasegawa}}\ and\ \bibinfo {author} {\bibfnamefont {Y.}~\bibnamefont
  {Tanimura}},\ }\bibfield  {title} {\enquote {\bibinfo {title} {Nonequilibrium
  molecular dynamics simulations with a backward-forward trajectories sampling
  for multidimensional infrared spectroscopy of molecular vibrational modes},}\
  }\href {https://doi.org/10.1063/1.2828189} {\bibfield  {journal} {\bibinfo
  {journal} {The Journal of Chemical Physics}\ }\textbf {\bibinfo {volume}
  {128}},\ \bibinfo {pages} {064511} (\bibinfo {year} {2008})}\BibitemShut
  {NoStop}%
\bibitem [{\citenamefont {Yagasaki}\ and\ \citenamefont
  {Saito}(2008)}]{YagasakiSaitoJCP20082DIR}%
  \BibitemOpen
  \bibfield  {author} {\bibinfo {author} {\bibfnamefont {T.}~\bibnamefont
  {Yagasaki}}\ and\ \bibinfo {author} {\bibfnamefont {S.}~\bibnamefont
  {Saito}},\ }\bibfield  {title} {\enquote {\bibinfo {title} {Ultrafast
  intermolecular dynamics of liquid water: A theoretical study on
  two-dimensional infrared spectroscopy},}\ }\href
  {https://doi.org/10.1063/1.2903470} {\bibfield  {journal} {\bibinfo
  {journal} {The Journal of Chemical Physics}\ }\textbf {\bibinfo {volume}
  {128}},\ \bibinfo {pages} {154521} (\bibinfo {year} {2008})},\ \Eprint
  {https://arxiv.org/abs/https://doi.org/10.1063/1.2903470}
  {https://doi.org/10.1063/1.2903470} \BibitemShut {NoStop}%
\bibitem [{\citenamefont {Liu}\ and\ \citenamefont
  {Liu}(2018)}]{JianLiu2018H2OMP}%
  \BibitemOpen
  \bibfield  {author} {\bibinfo {author} {\bibfnamefont {X.}~\bibnamefont
  {Liu}}\ and\ \bibinfo {author} {\bibfnamefont {J.}~\bibnamefont {Liu}},\
  }\bibfield  {title} {\enquote {\bibinfo {title} {Critical role of quantum
  dynamical effects in the \uppercase{R}aman spectroscopy of liquid water},}\
  }\href {https://doi.org/10.1080/00268976.2018.1434907} {\bibfield  {journal}
  {\bibinfo  {journal} {Molecular Physics}\ }\textbf {\bibinfo {volume}
  {116}},\ \bibinfo {pages} {755--779} (\bibinfo {year} {2018})},\ \Eprint
  {https://arxiv.org/abs/https://doi.org/10.1080/00268976.2018.1434907}
  {https://doi.org/10.1080/00268976.2018.1434907} \BibitemShut {NoStop}%
\bibitem [{\citenamefont {Hunter}, \citenamefont {Shakib},\ and\ \citenamefont
  {Paesani}(2018)}]{Paesan2018H2OCMD}%
  \BibitemOpen
  \bibfield  {author} {\bibinfo {author} {\bibfnamefont {K.~M.}\ \bibnamefont
  {Hunter}}, \bibinfo {author} {\bibfnamefont {F.~A.}\ \bibnamefont {Shakib}},\
  and\ \bibinfo {author} {\bibfnamefont {F.}~\bibnamefont {Paesani}},\
  }\bibfield  {title} {\enquote {\bibinfo {title} {Disentangling coupling
  effects in the infrared spectra of liquid water},}\ }\href
  {https://doi.org/10.1021/acs.jpcb.8b09910} {\bibfield  {journal} {\bibinfo
  {journal} {The Journal of Physical Chemistry B}\ }\textbf {\bibinfo {volume}
  {122}},\ \bibinfo {pages} {10754--10761} (\bibinfo {year} {2018})},\ \bibinfo
  {note} {pMID: 30403350},\ \Eprint
  {https://arxiv.org/abs/https://doi.org/10.1021/acs.jpcb.8b09910}
  {https://doi.org/10.1021/acs.jpcb.8b09910} \BibitemShut {NoStop}%
\bibitem [{\citenamefont {Trenins}, \citenamefont {Willatt},\ and\
  \citenamefont {Althorpe}(2019)}]{Althorpe2019CMD}%
  \BibitemOpen
  \bibfield  {author} {\bibinfo {author} {\bibfnamefont {G.}~\bibnamefont
  {Trenins}}, \bibinfo {author} {\bibfnamefont {M.~J.}\ \bibnamefont
  {Willatt}},\ and\ \bibinfo {author} {\bibfnamefont {S.~C.}\ \bibnamefont
  {Althorpe}},\ }\bibfield  {title} {\enquote {\bibinfo {title} {Path-integral
  dynamics of water using curvilinear centroids},}\ }\href
  {https://doi.org/10.1063/1.5100587} {\bibfield  {journal} {\bibinfo
  {journal} {The Journal of Chemical Physics}\ }\textbf {\bibinfo {volume}
  {151}},\ \bibinfo {pages} {054109} (\bibinfo {year} {2019})},\ \Eprint
  {https://arxiv.org/abs/https://doi.org/10.1063/1.5100587}
  {https://doi.org/10.1063/1.5100587} \BibitemShut {NoStop}%
\bibitem [{\citenamefont {Gottwald}, \citenamefont {Ivanov},\ and\
  \citenamefont {Kühn}(2015)}]{Kuehn2015JPCL}%
  \BibitemOpen
  \bibfield  {author} {\bibinfo {author} {\bibfnamefont {F.}~\bibnamefont
  {Gottwald}}, \bibinfo {author} {\bibfnamefont {S.~D.}\ \bibnamefont
  {Ivanov}},\ and\ \bibinfo {author} {\bibfnamefont {O.}~\bibnamefont
  {Kühn}},\ }\bibfield  {title} {\enquote {\bibinfo {title} {Applicability of
  the caldeira–leggett model to vibrational spectroscopy in solution},}\
  }\href {https://doi.org/10.1021/acs.jpclett.5b00718} {\bibfield  {journal}
  {\bibinfo  {journal} {The Journal of Physical Chemistry Letters}\ }\textbf
  {\bibinfo {volume} {6}},\ \bibinfo {pages} {2722--2727} (\bibinfo {year}
  {2015})}\BibitemShut {NoStop}%
\bibitem [{\citenamefont {Gottwald}\ \emph {et~al.}(2015)\citenamefont
  {Gottwald}, \citenamefont {Karsten}, \citenamefont {Ivanov},\ and\
  \citenamefont {Kühn}}]{Kuehn2015JCP}%
  \BibitemOpen
  \bibfield  {author} {\bibinfo {author} {\bibfnamefont {F.}~\bibnamefont
  {Gottwald}}, \bibinfo {author} {\bibfnamefont {S.}~\bibnamefont {Karsten}},
  \bibinfo {author} {\bibfnamefont {S.~D.}\ \bibnamefont {Ivanov}},\ and\
  \bibinfo {author} {\bibfnamefont {O.}~\bibnamefont {Kühn}},\ }\bibfield
  {title} {\enquote {\bibinfo {title} {Parametrizing linear generalized
  langevin dynamics from explicit molecular dynamics simulations},}\ }\href
  {https://doi.org/10.1063/1.4922941} {\bibfield  {journal} {\bibinfo
  {journal} {The Journal of Chemical Physics}\ }\textbf {\bibinfo {volume}
  {142}},\ \bibinfo {pages} {244110} (\bibinfo {year} {2015})}\BibitemShut
  {NoStop}%
\bibitem [{\citenamefont {Gottwald}, \citenamefont {Ivanov},\ and\
  \citenamefont {Kühn}(2016)}]{Kuehn2016JCP}%
  \BibitemOpen
  \bibfield  {author} {\bibinfo {author} {\bibfnamefont {F.}~\bibnamefont
  {Gottwald}}, \bibinfo {author} {\bibfnamefont {S.~D.}\ \bibnamefont
  {Ivanov}},\ and\ \bibinfo {author} {\bibfnamefont {O.}~\bibnamefont
  {Kühn}},\ }\bibfield  {title} {\enquote {\bibinfo {title} {Vibrational
  spectroscopy via the caldeira-leggett model with anharmonic system
  potentials},}\ }\href {https://doi.org/10.1063/1.4946872} {\bibfield
  {journal} {\bibinfo  {journal} {The Journal of Chemical Physics}\ }\textbf
  {\bibinfo {volume} {144}},\ \bibinfo {pages} {164102} (\bibinfo {year}
  {2016})}\BibitemShut {NoStop}%
\bibitem [{\citenamefont {Tanimura}\ and\ \citenamefont
  {Ishizaki}(2009)}]{TI09ACR}%
  \BibitemOpen
  \bibfield  {author} {\bibinfo {author} {\bibfnamefont {Y.}~\bibnamefont
  {Tanimura}}\ and\ \bibinfo {author} {\bibfnamefont {A.}~\bibnamefont
  {Ishizaki}},\ }\bibfield  {title} {\enquote {\bibinfo {title} {Modeling,
  calculating, and analyzing multidimensional vibrational spectroscopies},}\
  }\href {https://doi.org/10.1021/ar9000444} {\bibfield  {journal} {\bibinfo
  {journal} {Accounts of Chemical Research}\ }\textbf {\bibinfo {volume}
  {42}},\ \bibinfo {pages} {1270--1279} (\bibinfo {year} {2009})}\BibitemShut
  {NoStop}%
\bibitem [{\citenamefont {Yagasaki}\ and\ \citenamefont
  {Saito}(2011)}]{YagasakiSaitoJCP2011Relax}%
  \BibitemOpen
  \bibfield  {author} {\bibinfo {author} {\bibfnamefont {T.}~\bibnamefont
  {Yagasaki}}\ and\ \bibinfo {author} {\bibfnamefont {S.}~\bibnamefont
  {Saito}},\ }\bibfield  {title} {\enquote {\bibinfo {title} {A novel method
  for analyzing energy relaxation in condensed phases using nonequilibrium
  molecular dynamics simulations: Application to the energy relaxation of
  intermolecular motions in liquid water},}\ }\href
  {https://doi.org/10.1063/1.3587105} {\bibfield  {journal} {\bibinfo
  {journal} {The Journal of Chemical Physics}\ }\textbf {\bibinfo {volume}
  {134}},\ \bibinfo {pages} {184503} (\bibinfo {year} {2011})},\ \Eprint
  {https://arxiv.org/abs/https://doi.org/10.1063/1.3587105}
  {https://doi.org/10.1063/1.3587105} \BibitemShut {NoStop}%
\bibitem [{\citenamefont {Okumura}\ and\ \citenamefont
  {Tanimura}(1997{\natexlab{a}})}]{OT97PRE}%
  \BibitemOpen
  \bibfield  {author} {\bibinfo {author} {\bibfnamefont {K.}~\bibnamefont
  {Okumura}}\ and\ \bibinfo {author} {\bibfnamefont {Y.}~\bibnamefont
  {Tanimura}},\ }\bibfield  {title} {\enquote {\bibinfo {title} {Two-time
  correlation functions of a harmonic system nonbilinearly coupled to a heat
  bath: Spontaneous \uppercase{R}aman spectroscopy},}\ }\href
  {https://doi.org/10.1103/PhysRevE.56.2747} {\bibfield  {journal} {\bibinfo
  {journal} {Phys. Rev. E}\ }\textbf {\bibinfo {volume} {56}},\ \bibinfo
  {pages} {2747--2750} (\bibinfo {year} {1997}{\natexlab{a}})}\BibitemShut
  {NoStop}%
\bibitem [{\citenamefont {Ishizaki}\ and\ \citenamefont
  {Tanimura}(2007)}]{IT07JPCA}%
  \BibitemOpen
  \bibfield  {author} {\bibinfo {author} {\bibfnamefont {A.}~\bibnamefont
  {Ishizaki}}\ and\ \bibinfo {author} {\bibfnamefont {Y.}~\bibnamefont
  {Tanimura}},\ }\bibfield  {title} {\enquote {\bibinfo {title} {Dynamics of a
  multimode system coupled to multiple heat baths probed by two-dimensional
  infrared spectroscopy},}\ }\href {https://doi.org/10.1021/jp072880a}
  {\bibfield  {journal} {\bibinfo  {journal} {The Journal of Physical Chemistry
  A}\ }\textbf {\bibinfo {volume} {111}},\ \bibinfo {pages} {9269--9276}
  (\bibinfo {year} {2007})}\BibitemShut {NoStop}%
\bibitem [{\citenamefont {Tanimura}(2006)}]{T06JPSJ}%
  \BibitemOpen
  \bibfield  {author} {\bibinfo {author} {\bibfnamefont {Y.}~\bibnamefont
  {Tanimura}},\ }\bibfield  {title} {\enquote {\bibinfo {title} {Stochastic
  \uppercase{L}iouville, \uppercase{L}angevin,
  \uppercase{F}okker-\uppercase{P}lanck, and master equation qpproaches to
  quantum dissipative systems},}\ }\href
  {https://doi.org/10.1143/JPSJ.75.082001} {\bibfield  {journal} {\bibinfo
  {journal} {Journal of the Physical Society of Japan}\ }\textbf {\bibinfo
  {volume} {75}},\ \bibinfo {pages} {082001} (\bibinfo {year}
  {2006})}\BibitemShut {NoStop}%
\bibitem [{\citenamefont {Tanimura}(2020)}]{T20JCP}%
  \BibitemOpen
  \bibfield  {author} {\bibinfo {author} {\bibfnamefont {Y.}~\bibnamefont
  {Tanimura}},\ }\bibfield  {title} {\enquote {\bibinfo {title} {Numerically
  "exact" approach to open quantum dynamics: The hierarchical equations of
  motion (\uppercase{HEOM})},}\ }\href {https://doi.org/10.1063/5.0011599}
  {\bibfield  {journal} {\bibinfo  {journal} {The Journal of Chemical Physics}\
  }\textbf {\bibinfo {volume} {153}},\ \bibinfo {pages} {020901} (\bibinfo
  {year} {2020})}\BibitemShut {NoStop}%
\bibitem [{\citenamefont {Ikeda}, \citenamefont {Ito},\ and\ \citenamefont
  {Tanimura}(2015)}]{IIT15JCP}%
  \BibitemOpen
  \bibfield  {author} {\bibinfo {author} {\bibfnamefont {T.}~\bibnamefont
  {Ikeda}}, \bibinfo {author} {\bibfnamefont {H.}~\bibnamefont {Ito}},\ and\
  \bibinfo {author} {\bibfnamefont {Y.}~\bibnamefont {Tanimura}},\ }\bibfield
  {title} {\enquote {\bibinfo {title} {Analysis of 2d
  \uppercase{TH}z-\uppercase{R}aman spectroscopy using a
  non-\uppercase{M}arkovian \uppercase{B}rownian oscillator model with
  nonlinear system-bath interactions},}\ }\href
  {https://doi.org/10.1063/1.4917033} {\bibfield  {journal} {\bibinfo
  {journal} {The Journal of Chemical Physics}\ }\textbf {\bibinfo {volume}
  {142}},\ \bibinfo {pages} {212421} (\bibinfo {year} {2015})}\BibitemShut
  {NoStop}%
\bibitem [{\citenamefont {Ito}\ and\ \citenamefont {Tanimura}(2016)}]{IT16JCP}%
  \BibitemOpen
  \bibfield  {author} {\bibinfo {author} {\bibfnamefont {H.}~\bibnamefont
  {Ito}}\ and\ \bibinfo {author} {\bibfnamefont {Y.}~\bibnamefont {Tanimura}},\
  }\bibfield  {title} {\enquote {\bibinfo {title} {Simulating two-dimensional
  infrared-\uppercase{R}aman and \uppercase{R}aman spectroscopies for
  intermolecular and intramolecular modes of liquid water},}\ }\href
  {https://doi.org/10.1063/1.4941842} {\bibfield  {journal} {\bibinfo
  {journal} {The Journal of Chemical Physics}\ }\textbf {\bibinfo {volume}
  {144}},\ \bibinfo {pages} {074201} (\bibinfo {year} {2016})}\BibitemShut
  {NoStop}%
\bibitem [{\citenamefont {Tanimura}\ and\ \citenamefont
  {Wolynes}(1991)}]{TW91PRA}%
  \BibitemOpen
  \bibfield  {author} {\bibinfo {author} {\bibfnamefont {Y.}~\bibnamefont
  {Tanimura}}\ and\ \bibinfo {author} {\bibfnamefont {P.~G.}\ \bibnamefont
  {Wolynes}},\ }\bibfield  {title} {\enquote {\bibinfo {title} {Quantum and
  classical \uppercase{F}okker-\uppercase{P}lanck equations for a
  \uppercase{G}aussian-\uppercase{M}arkovian noise bath},}\ }\href
  {https://doi.org/10.1103/PhysRevA.43.4131} {\bibfield  {journal} {\bibinfo
  {journal} {Phys. Rev. A}\ }\textbf {\bibinfo {volume} {43}},\ \bibinfo
  {pages} {4131--4142} (\bibinfo {year} {1991})}\BibitemShut {NoStop}%
\bibitem [{\citenamefont {Tanimura}\ and\ \citenamefont
  {Kubo}(1989)}]{TK89JPSJ1}%
  \BibitemOpen
  \bibfield  {author} {\bibinfo {author} {\bibfnamefont {Y.}~\bibnamefont
  {Tanimura}}\ and\ \bibinfo {author} {\bibfnamefont {R.}~\bibnamefont
  {Kubo}},\ }\bibfield  {title} {\enquote {\bibinfo {title} {Time evolution of
  a quantum system in contact with a nearly
  \uppercase{G}aussian-\uppercase{M}arkoffian noise bath},}\ }\href
  {https://doi.org/10.1143/JPSJ.58.101} {\bibfield  {journal} {\bibinfo
  {journal} {Journal of the Physical Society of Japan}\ }\textbf {\bibinfo
  {volume} {58}},\ \bibinfo {pages} {101--114} (\bibinfo {year}
  {1989})}\BibitemShut {NoStop}%
\bibitem [{\citenamefont {Tanimura}\ and\ \citenamefont
  {Steffen}(2000)}]{TS20JPSJ}%
  \BibitemOpen
  \bibfield  {author} {\bibinfo {author} {\bibfnamefont {Y.}~\bibnamefont
  {Tanimura}}\ and\ \bibinfo {author} {\bibfnamefont {T.}~\bibnamefont
  {Steffen}},\ }\bibfield  {title} {\enquote {\bibinfo {title} {Two-dimensional
  spectroscopy for harmonic vibrational modes with nonlinear system-bath
  interactions.ii. \uppercase{G}aussian-\uppercase{M}arkovian case},}\ }\href
  {https://doi.org/10.1143/JPSJ.69.4095} {\bibfield  {journal} {\bibinfo
  {journal} {Journal of the Physical Society of Japan}\ }\textbf {\bibinfo
  {volume} {69}},\ \bibinfo {pages} {4095--4106} (\bibinfo {year}
  {2000})}\BibitemShut {NoStop}%
\bibitem [{\citenamefont {Kato}\ and\ \citenamefont
  {Tanimura}(2002)}]{KT02JCP1}%
  \BibitemOpen
  \bibfield  {author} {\bibinfo {author} {\bibfnamefont {T.}~\bibnamefont
  {Kato}}\ and\ \bibinfo {author} {\bibfnamefont {Y.}~\bibnamefont
  {Tanimura}},\ }\bibfield  {title} {\enquote {\bibinfo {title} {Vibrational
  spectroscopy of a harmonic oscillator system nonlinearly coupled to a heat
  bath},}\ }\href {https://doi.org/10.1063/1.1503778} {\bibfield  {journal}
  {\bibinfo  {journal} {The Journal of Chemical Physics}\ }\textbf {\bibinfo
  {volume} {117}},\ \bibinfo {pages} {6221--6234} (\bibinfo {year}
  {2002})}\BibitemShut {NoStop}%
\bibitem [{\citenamefont {Kato}\ and\ \citenamefont
  {Tanimura}(2004)}]{KT04JCP}%
  \BibitemOpen
  \bibfield  {author} {\bibinfo {author} {\bibfnamefont {T.}~\bibnamefont
  {Kato}}\ and\ \bibinfo {author} {\bibfnamefont {Y.}~\bibnamefont
  {Tanimura}},\ }\bibfield  {title} {\enquote {\bibinfo {title}
  {Two-dimensional \uppercase{R}aman and infrared vibrational spectroscopy for
  a harmonic oscillator system nonlinearly coupled with a colored noise
  bath},}\ }\href {https://doi.org/10.1063/1.1629272} {\bibfield  {journal}
  {\bibinfo  {journal} {The Journal of Chemical Physics}\ }\textbf {\bibinfo
  {volume} {120}},\ \bibinfo {pages} {260--271} (\bibinfo {year}
  {2004})}\BibitemShut {NoStop}%
\bibitem [{\citenamefont {Tanimura}(2014)}]{T14JCP}%
  \BibitemOpen
  \bibfield  {author} {\bibinfo {author} {\bibfnamefont {Y.}~\bibnamefont
  {Tanimura}},\ }\bibfield  {title} {\enquote {\bibinfo {title} {Reduced
  hierarchical equations of motion in real and imaginary time: Correlated
  initial states and thermodynamic quantities},}\ }\href
  {https://doi.org/10.1063/1.4890441} {\bibfield  {journal} {\bibinfo
  {journal} {The Journal of Chemical Physics}\ }\textbf {\bibinfo {volume}
  {141}},\ \bibinfo {pages} {044114} (\bibinfo {year} {2014})}\BibitemShut
  {NoStop}%
\bibitem [{\citenamefont {Steinel}\ \emph {et~al.}(2004)\citenamefont
  {Steinel}, \citenamefont {Asbury}, \citenamefont {Corcelli}, \citenamefont
  {Lawrence}, \citenamefont {Skinner},\ and\ \citenamefont
  {Fayer}}]{SkinnerCPL2004}%
  \BibitemOpen
  \bibfield  {author} {\bibinfo {author} {\bibfnamefont {T.}~\bibnamefont
  {Steinel}}, \bibinfo {author} {\bibfnamefont {J.~B.}\ \bibnamefont {Asbury}},
  \bibinfo {author} {\bibfnamefont {S.}~\bibnamefont {Corcelli}}, \bibinfo
  {author} {\bibfnamefont {C.}~\bibnamefont {Lawrence}}, \bibinfo {author}
  {\bibfnamefont {J.}~\bibnamefont {Skinner}},\ and\ \bibinfo {author}
  {\bibfnamefont {M.}~\bibnamefont {Fayer}},\ }\bibfield  {title} {\enquote
  {\bibinfo {title} {Water dynamics: dependence on local structure probed with
  vibrational echo correlation spectroscopy},}\ }\href
  {https://doi.org/https://doi.org/10.1016/j.cplett.2004.01.042} {\bibfield
  {journal} {\bibinfo  {journal} {Chemical Physics Letters}\ }\textbf {\bibinfo
  {volume} {386}},\ \bibinfo {pages} {295--300} (\bibinfo {year}
  {2004})}\BibitemShut {NoStop}%
\bibitem [{\citenamefont {Light}, \citenamefont {Hamilton},\ and\ \citenamefont
  {Lill}(1985)}]{light1985}%
  \BibitemOpen
  \bibfield  {author} {\bibinfo {author} {\bibfnamefont {J.~C.}\ \bibnamefont
  {Light}}, \bibinfo {author} {\bibfnamefont {I.~P.}\ \bibnamefont
  {Hamilton}},\ and\ \bibinfo {author} {\bibfnamefont {J.~V.}\ \bibnamefont
  {Lill}},\ }\bibfield  {title} {\enquote {\bibinfo {title} {Generalized
  discrete variable approximation in quantum mechanics},}\ }\href
  {https://doi.org/10.1063/1.448462} {\bibfield  {journal} {\bibinfo  {journal}
  {J. Chem. Phys.}\ }\textbf {\bibinfo {volume} {82}},\ \bibinfo {pages}
  {1400--1409} (\bibinfo {year} {1985})}\BibitemShut {NoStop}%
\bibitem [{\citenamefont {Baye}\ and\ \citenamefont {Heenen}(1986)}]{baye1986}%
  \BibitemOpen
  \bibfield  {author} {\bibinfo {author} {\bibfnamefont {D.}~\bibnamefont
  {Baye}}\ and\ \bibinfo {author} {\bibfnamefont {P.-H.}\ \bibnamefont
  {Heenen}},\ }\bibfield  {title} {\enquote {\bibinfo {title} {Generalised
  meshes for quantum mechanical problems},}\ }\href
  {https://doi.org/10.1088/0305-4470/19/11/013} {\bibfield  {journal} {\bibinfo
   {journal} {J. Phys. A: Math. Gen.}\ }\textbf {\bibinfo {volume} {19}},\
  \bibinfo {pages} {2041--2059} (\bibinfo {year} {1986})}\BibitemShut {NoStop}%
\bibitem [{\citenamefont {Baye}(2015)}]{baye2015}%
  \BibitemOpen
  \bibfield  {author} {\bibinfo {author} {\bibfnamefont {D.}~\bibnamefont
  {Baye}},\ }\bibfield  {title} {\enquote {\bibinfo {title} {The
  {{Lagrange-mesh}} method},}\ }\href
  {https://doi.org/10.1016/j.physrep.2014.11.006} {\bibfield  {journal}
  {\bibinfo  {journal} {Physics Reports}\ }\bibinfo {series} {The
  {{Lagrange-mesh}} Method},\ \textbf {\bibinfo {volume} {565}},\ \bibinfo
  {pages} {1--107} (\bibinfo {year} {2015})}\BibitemShut {NoStop}%
\bibitem [{\citenamefont {Erpenbeck}\ and\ \citenamefont
  {Thoss}(2019)}]{erpenbeck2019}%
  \BibitemOpen
  \bibfield  {author} {\bibinfo {author} {\bibfnamefont {A.}~\bibnamefont
  {Erpenbeck}}\ and\ \bibinfo {author} {\bibfnamefont {M.}~\bibnamefont
  {Thoss}},\ }\bibfield  {title} {\enquote {\bibinfo {title} {Hierarchical
  quantum master equation approach to vibronic reaction dynamics at metal
  surfaces},}\ }\href {https://doi.org/10.1063/1.5128206} {\bibfield  {journal}
  {\bibinfo  {journal} {J. Chem. Phys.}\ }\textbf {\bibinfo {volume} {151}},\
  \bibinfo {pages} {191101} (\bibinfo {year} {2019})}\BibitemShut {NoStop}%
\bibitem [{\citenamefont {Tanimura}(2015)}]{T15JCP}%
  \BibitemOpen
  \bibfield  {author} {\bibinfo {author} {\bibfnamefont {Y.}~\bibnamefont
  {Tanimura}},\ }\bibfield  {title} {\enquote {\bibinfo {title} {Real-time and
  imaginary-time quantum hierarchal \uppercase{F}okker-\uppercase{P}lanck
  equations},}\ }\href {https://doi.org/10.1063/1.4916647} {\bibfield
  {journal} {\bibinfo  {journal} {The Journal of Chemical Physics}\ }\textbf
  {\bibinfo {volume} {142}},\ \bibinfo {pages} {144110} (\bibinfo {year}
  {2015})}\BibitemShut {NoStop}%
\bibitem [{\citenamefont {Blackmore}\ and\ \citenamefont
  {Shizgal}(1985)}]{blackmore1985a}%
  \BibitemOpen
  \bibfield  {author} {\bibinfo {author} {\bibfnamefont {R.}~\bibnamefont
  {Blackmore}}\ and\ \bibinfo {author} {\bibfnamefont {B.}~\bibnamefont
  {Shizgal}},\ }\bibfield  {title} {\enquote {\bibinfo {title} {A solution of
  {{Kramers}} equation for the isomerization of n-butane in {{CCl4}}},}\ }\href
  {https://doi.org/10.1063/1.449247} {\bibfield  {journal} {\bibinfo  {journal}
  {J. Chem. Phys.}\ }\textbf {\bibinfo {volume} {83}},\ \bibinfo {pages}
  {2934--2941} (\bibinfo {year} {1985})}\BibitemShut {NoStop}%
\bibitem [{\citenamefont {Schwartz}(1985)}]{schwartz1985}%
  \BibitemOpen
  \bibfield  {author} {\bibinfo {author} {\bibfnamefont {C.}~\bibnamefont
  {Schwartz}},\ }\bibfield  {title} {\enquote {\bibinfo {title} {High-accuracy
  approximation techniques for analytic functions},}\ }\href
  {https://doi.org/10.1063/1.526624} {\bibfield  {journal} {\bibinfo  {journal}
  {J. Math. Phys.}\ }\textbf {\bibinfo {volume} {26}},\ \bibinfo {pages}
  {411--415} (\bibinfo {year} {1985})}\BibitemShut {NoStop}%
\bibitem [{\citenamefont {Iwamoto}\ and\ \citenamefont
  {Tanimura}(2021)}]{IT21JCEL}%
  \BibitemOpen
  \bibfield  {author} {\bibinfo {author} {\bibfnamefont {Y.}~\bibnamefont
  {Iwamoto}}\ and\ \bibinfo {author} {\bibfnamefont {Y.}~\bibnamefont
  {Tanimura}},\ }\bibfield  {title} {\enquote {\bibinfo {title} {Open quantum
  dynamics theory on the basis of periodical system-bath model for discrete
  \uppercase{W}igner function},}\ }\href
  {https://doi.org/110.1007/s10825-021-01754-z} {\bibfield  {journal} {\bibinfo
   {journal} {Journal of Computational Electronics}\ }\textbf {\bibinfo
  {volume} {20}},\ \bibinfo {pages} {2091–2103} (\bibinfo {year}
  {2021})}\BibitemShut {NoStop}%
\bibitem [{\citenamefont {Yan}(2017)}]{yan2017cjcp}%
  \BibitemOpen
  \bibfield  {author} {\bibinfo {author} {\bibfnamefont {Y.-a.}\ \bibnamefont
  {Yan}},\ }\bibfield  {title} {\enquote {\bibinfo {title} {{Low-Storage
  Runge-Kutta Method for Simulating Time-Dependent Quantum Dynamics}},}\ }\href
  {https://doi.org/10.1063/1674-0068/30/cjcp1703025} {\bibfield  {journal}
  {\bibinfo  {journal} {Chin. J. Chem. Phys.}\ }\textbf {\bibinfo {volume}
  {30}},\ \bibinfo {pages} {277} (\bibinfo {year} {2017})}\BibitemShut
  {NoStop}%
\bibitem [{\citenamefont {Ikeda}\ and\ \citenamefont
  {Tanimura}(2019)}]{IT19JCTC}%
  \BibitemOpen
  \bibfield  {author} {\bibinfo {author} {\bibfnamefont {T.}~\bibnamefont
  {Ikeda}}\ and\ \bibinfo {author} {\bibfnamefont {Y.}~\bibnamefont
  {Tanimura}},\ }\bibfield  {title} {\enquote {\bibinfo {title}
  {Low-temperature quantum \uppercase{F}okker-\uppercase{P}lanck and
  \uppercase{S}moluchowski equations and their extension to multistate
  systems},}\ }\href {https://doi.org/10.1021/acs.jctc.8b01195} {\bibfield
  {journal} {\bibinfo  {journal} {Journal of Chemical Theory and Computation}\
  }\textbf {\bibinfo {volume} {15}},\ \bibinfo {pages} {2517--2534} (\bibinfo
  {year} {2019})}\BibitemShut {NoStop}%
\bibitem [{\citenamefont {Okumura}\ and\ \citenamefont
  {Tanimura}(1997{\natexlab{b}})}]{OT97CPL2}%
  \BibitemOpen
  \bibfield  {author} {\bibinfo {author} {\bibfnamefont {K.}~\bibnamefont
  {Okumura}}\ and\ \bibinfo {author} {\bibfnamefont {Y.}~\bibnamefont
  {Tanimura}},\ }\bibfield  {title} {\enquote {\bibinfo {title} {Sensitivity of
  two-dimensional fifth-order \uppercase{R}aman response to the mechanism of
  vibrational mode-mode coupling in liquid molecules},}\ }\href
  {https://doi.org/https://doi.org/10.1016/S0009-2614(97)00942-1} {\bibfield
  {journal} {\bibinfo  {journal} {Chemical Physics Letters}\ }\textbf {\bibinfo
  {volume} {278}},\ \bibinfo {pages} {175--183} (\bibinfo {year}
  {1997}{\natexlab{b}})}\BibitemShut {NoStop}%
\bibitem [{\citenamefont {Suzuki}\ and\ \citenamefont
  {Tanimura}(2002)}]{ST02CPL}%
  \BibitemOpen
  \bibfield  {author} {\bibinfo {author} {\bibfnamefont {Y.}~\bibnamefont
  {Suzuki}}\ and\ \bibinfo {author} {\bibfnamefont {Y.}~\bibnamefont
  {Tanimura}},\ }\bibfield  {title} {\enquote {\bibinfo {title} {Probing a
  colored-noise induced peak of a strongly damped \uppercase{B}rownian system
  by one- and two-dimensional spectroscopy},}\ }\href
  {https://doi.org/https://doi.org/10.1016/S0009-2614(02)00563-8} {\bibfield
  {journal} {\bibinfo  {journal} {Chemical Physics Letters}\ }\textbf {\bibinfo
  {volume} {358}},\ \bibinfo {pages} {51--56} (\bibinfo {year}
  {2002})}\BibitemShut {NoStop}%
\bibitem [{\citenamefont {Ueno}\ and\ \citenamefont
  {Tanimura}(2020)}]{UT20JCTC}%
  \BibitemOpen
  \bibfield  {author} {\bibinfo {author} {\bibfnamefont {S.}~\bibnamefont
  {Ueno}}\ and\ \bibinfo {author} {\bibfnamefont {Y.}~\bibnamefont
  {Tanimura}},\ }\bibfield  {title} {\enquote {\bibinfo {title} {Modeling
  intermolecular and intramolecular modes of liquid water using multiple heat
  baths: Machine learning approach},}\ }\href
  {https://doi.org/10.1021/acs.jctc.9b01288} {\bibfield  {journal} {\bibinfo
  {journal} {Journal of Chemical Theory and Computation}\ }\textbf {\bibinfo
  {volume} {16}},\ \bibinfo {pages} {2099--2108} (\bibinfo {year}
  {2020})}\BibitemShut {NoStop}%
\bibitem [{\citenamefont {Tian}\ \emph {et~al.}(2010)\citenamefont {Tian},
  \citenamefont {Ding}, \citenamefont {Xu},\ and\ \citenamefont
  {Yan}}]{YanPade2010}%
  \BibitemOpen
  \bibfield  {author} {\bibinfo {author} {\bibfnamefont {B.-L.}\ \bibnamefont
  {Tian}}, \bibinfo {author} {\bibfnamefont {J.-J.}\ \bibnamefont {Ding}},
  \bibinfo {author} {\bibfnamefont {R.-X.}\ \bibnamefont {Xu}},\ and\ \bibinfo
  {author} {\bibfnamefont {Y.}~\bibnamefont {Yan}},\ }\bibfield  {title}
  {\enquote {\bibinfo {title} {Biexponential theory of drude dissipation via
  hierarchical quantum master equation},}\ }\href
  {https://doi.org/10.1063/1.3491270} {\bibfield  {journal} {\bibinfo
  {journal} {The Journal of Chemical Physics}\ }\textbf {\bibinfo {volume}
  {133}},\ \bibinfo {pages} {114112} (\bibinfo {year} {2010})},\ \Eprint
  {https://arxiv.org/abs/https://doi.org/10.1063/1.3491270}
  {https://doi.org/10.1063/1.3491270} \BibitemShut {NoStop}%
\bibitem [{\citenamefont {Xu}\ and\ \citenamefont {Jiang}(2013)}]{xu2013}%
  \BibitemOpen
  \bibfield  {author} {\bibinfo {author} {\bibfnamefont {K.}~\bibnamefont
  {Xu}}\ and\ \bibinfo {author} {\bibfnamefont {S.}~\bibnamefont {Jiang}},\
  }\bibfield  {title} {\enquote {\bibinfo {title} {A {{Bootstrap Method}} for
  {{Sum-of-Poles Approximations}}},}\ }\href
  {https://doi.org/10.1007/s10915-012-9620-9} {\bibfield  {journal} {\bibinfo
  {journal} {J Sci Comput}\ }\textbf {\bibinfo {volume} {55}},\ \bibinfo
  {pages} {16--39} (\bibinfo {year} {2013})}\BibitemShut {NoStop}%
\bibitem [{\citenamefont {Ikeno}(2018)}]{ikeno2018}%
  \BibitemOpen
  \bibfield  {author} {\bibinfo {author} {\bibfnamefont {H.}~\bibnamefont
  {Ikeno}},\ }\bibfield  {title} {\enquote {\bibinfo {title} {Mxpfit: {{A}}
  library for finding optimal multi-exponential approximations},}\ }\href
  {https://doi.org/10.1016/j.cpc.2018.04.015} {\bibfield  {journal} {\bibinfo
  {journal} {Computer Physics Communications}\ }\textbf {\bibinfo {volume}
  {230}},\ \bibinfo {pages} {135--144} (\bibinfo {year} {2018})}\BibitemShut
  {NoStop}%
\bibitem [{\citenamefont {Fay}(2022)}]{fay2022b}%
  \BibitemOpen
  \bibfield  {author} {\bibinfo {author} {\bibfnamefont {T.~P.}\ \bibnamefont
  {Fay}},\ }\bibfield  {title} {\enquote {\bibinfo {title} {A simple improved
  low temperature correction for the hierarchical equations of motion},}\
  }\href {https://doi.org/10.1063/5.0100365} {\bibfield  {journal} {\bibinfo
  {journal} {J. Chem. Phys.}\ }\textbf {\bibinfo {volume} {157}},\ \bibinfo
  {pages} {054108} (\bibinfo {year} {2022})}\BibitemShut {NoStop}%
\bibitem [{\citenamefont {Tanimura}(1998)}]{T98CP}%
  \BibitemOpen
  \bibfield  {author} {\bibinfo {author} {\bibfnamefont {Y.}~\bibnamefont
  {Tanimura}},\ }\bibfield  {title} {\enquote {\bibinfo {title} {Fifth-order
  two-dimensional vibrational spectroscopy of a \uppercase{M}orse potential
  system in condensed phases},}\ }\href
  {https://doi.org/https://doi.org/10.1016/S0301-0104(98)00010-X} {\bibfield
  {journal} {\bibinfo  {journal} {Chemical Physics}\ }\textbf {\bibinfo
  {volume} {233}},\ \bibinfo {pages} {217--229} (\bibinfo {year}
  {1998})}\BibitemShut {NoStop}%
\bibitem [{\citenamefont {Dunn}, \citenamefont {Tempelaar},\ and\ \citenamefont
  {Reichman}(2019)}]{dunn2019}%
  \BibitemOpen
  \bibfield  {author} {\bibinfo {author} {\bibfnamefont {I.~S.}\ \bibnamefont
  {Dunn}}, \bibinfo {author} {\bibfnamefont {R.}~\bibnamefont {Tempelaar}},\
  and\ \bibinfo {author} {\bibfnamefont {D.~R.}\ \bibnamefont {Reichman}},\
  }\bibfield  {title} {\enquote {\bibinfo {title} {Removing instabilities in
  the hierarchical equations of motion: {{Exact}} and approximate projection
  approaches},}\ }\href {https://doi.org/10.1063/1.5092616} {\bibfield
  {journal} {\bibinfo  {journal} {J. Chem. Phys.}\ }\textbf {\bibinfo {volume}
  {150}},\ \bibinfo {pages} {184109} (\bibinfo {year} {2019})}\BibitemShut
  {NoStop}%
\bibitem [{\citenamefont {Chen}\ \emph {et~al.}(2022)\citenamefont {Chen},
  \citenamefont {Wang}, \citenamefont {Zheng}, \citenamefont {Xu},\ and\
  \citenamefont {Yan}}]{chen2022}%
  \BibitemOpen
  \bibfield  {author} {\bibinfo {author} {\bibfnamefont {Z.-H.}\ \bibnamefont
  {Chen}}, \bibinfo {author} {\bibfnamefont {Y.}~\bibnamefont {Wang}}, \bibinfo
  {author} {\bibfnamefont {X.}~\bibnamefont {Zheng}}, \bibinfo {author}
  {\bibfnamefont {R.-X.}\ \bibnamefont {Xu}},\ and\ \bibinfo {author}
  {\bibfnamefont {Y.}~\bibnamefont {Yan}},\ }\bibfield  {title} {\enquote
  {\bibinfo {title} {Universal time-domain {{Prony}} fitting decomposition for
  optimized hierarchical quantum master equations},}\ }\href
  {https://doi.org/10.1063/5.0095961} {\bibfield  {journal} {\bibinfo
  {journal} {J. Chem. Phys.}\ }\textbf {\bibinfo {volume} {156}},\ \bibinfo
  {pages} {221102} (\bibinfo {year} {2022})}\BibitemShut {NoStop}%
\end{thebibliography}%

\end{document}